\documentclass[12pt]{iopart}

\usepackage{amssymb}   
\usepackage[latin1]{inputenc} 
\usepackage[dvipdf]{epsfig}   
\usepackage{graphicx}
\usepackage{psfrag}
\usepackage[dvipdf]{epsfig}
\usepackage{hyperref}
\usepackage{tabularx}
\usepackage{doipubmed}
\usepackage{booktabs}

\begin{document}

\title[Exemplifying the equivalence of discrete quantum walk models]
{Unveiling and exemplifying the unitary equivalence of discrete time
quantum walk models}

\author{B F Venancio$^1$, F M Andrade$^2$, M G E da Luz$^1$}
\address{$^1$ Departamento de F\'{\i}sica,
Universidade Federal do Paran\'a,
C.P. 19044, 81531-980 Curitiba-PR, Brazil}
\address{$^2$ Departamento de Matem\'atica e Estat\'{\i}stica,
Universidade Estadual de Ponta Grossa,
84030-900 Ponta Grossa-PR, Brazil}
\ead{bfvenancio@fisica.ufpr.br, fmandrade@uepg.br, \\ luz@fisica.ufpr.br
(corresponding author)}

\begin{abstract}
The two major discrete time formulations for quantum walks, coined and 
scattering, are unitarily equivalent for arbitrary position dependent 
transition amplitudes and any topology (PRA {\bf 80}, 052301 (2009)).
Although the proof explicit describes the mapping obtention, its high
technicality may hinder relevant physical aspects involved in the 
equivalence.
Discussing concrete examples -- the most general constructions for the
line, square and honeycomb lattices -- here we unveil the similarities 
and differences of these two versions of quantum walks.
We moreover show how to derive the dynamics of one from the other by 
means of proper projections.
We perform calculations for different probability amplitudes like, Hadamard, 
Grover, Discrete Fourier Transform and the uncommon in the area (but 
interesting) Discrete Hartley Transform, comparing the evolutions.
Our study illustrates the models interplay, an important issue for 
implementations and applications of such systems.

\end{abstract}


\submitto{\JPA}

\maketitle

\section{Introduction}

Since their introduction 
\cite{Phys.Rev.A.48.1687.1993,jsp.1996.85.551,watrous-2001}, 
quantum walks (QWs), a paradigmatic and relatively simple class of quantum 
systems \cite{kempe1}, have found many applications in different areas of 
science (see, e.g., the recent review in \cite{venegas}).
At a first glance, such usefulness may be attributed to QWs distinct possible 
formulations.
For instance, although in all cases the dynamics take place in discrete
spatial structures, graphs (or lattices), time can be either a continuous 
\cite{ctv} or a discrete variable.
Moreover, in discrete time versions, the evolution can be dictated by inner 
states called ``coins'' \cite{tregenna1} (coined QWs, CQWs) or resulting 
from scattering-like processes \cite{hillery-2003,hillery-2004,hillery-2007} 
(scattering QWs, SQWs).

Regarding usages for QWs, some aspects of the general searching 
problem \cite{search} seem to be very appropriate for SQWs \cite{search1}, 
including identification and comparison of parts of a graph 
\cite{hillery-search,hillery-search-1}.
Also, SQWs are frequently considered in the study of scattering in 
semi-infinite graphs \cite{hillery-2004,hillery-2007,kosik}, a relevant 
configuration to implement universal quantum computation through QWs 
\cite{childs1}. 
On the other hand, the literature proposes a much larger number of 
applications for CQWs \cite{venegas,karski,leung,kempe-ptrf}. 
Examples are: transport in biological systems \cite{biology}
(in a continuous time context, see \cite{biology1}); 
Bose-Einstein condensates redistribution \cite{chandrashekar1}; 
quantum phase transition in optical lattices \cite{chandrashekar2}; 
decoherence processes \cite{ampadu}; and even quantum games \cite{game}.
In particular, CQWs are widely discussed in quantum computation as a tool
for the development of quantum algorithms \cite{qw-qcomput}.

As it concerns implementations, distinct experimental setups based on 
distinct physical phenomena can be used to actually build QWs.
For CQWs, protocols based on trapped atoms in optical lattices \cite{dur}, 
quantum dots \cite{hoogdalem}, photons orbital angular momentum 
\cite{zou}, and QED cavities \cite{tiegang}, to cite just a few, have been 
devised.
Concrete lab realizations were constructed with waveguide lattices 
\cite{crespi} (refers to \cite{perets} for waveguides in the continuous 
time case), passive optical elements \cite{schreiber}, and liquid-state 
nuclear magnetic resonance \cite{ryan}, again only mentioning few examples
(an overview is given in \cite{matjeschk}).
By their turn, SQWs could be associated to optical networks 
\cite{jex-beam,torma} and 
eventually may be ensemble just with linear optical elements 
\cite{search1,hillery-search}, similarly as done in \cite{do} to fabricate 
a quantum version of Galton's quincunx (a classical mechanical machine 
which at certain locations ``chooses'' -- with 50\%:50\% probability -- 
between two directions to go for a traveling ball).

The above mentioned models distinctions for applications (and in a less 
extend for implementations) are, however, due to practical and operation 
instead to fundamental reasons.
Indeed, continuous time and coined QWs are closely related, since they lead 
to a same dynamics in proper limits \cite{strauch,childs-limit} (although 
the limits may not be so direct to achieve \cite{childs-limit}).
Furthermore, it has been rigorously proven that CQWs and SQWs are 
unconditionally (i.e., for arbitrary topologies and spatially dependent 
transition probabilities) unitary equivalent \cite{andrade1}.

These kindred relations open important perspectives in the employment of 
QWs, specially for the discrete time versions. 
They should be equally appropriate in any application and in 
principle (see comments in Sec. 4), a same physical implementation would be 
capable to simulate both CQWs and SQWs.
But to benefit from such connections, it is necessary a clear understanding 
of similarities and differences of the two system versions as well as the 
correct way to obtain one from another.
The proof in \cite{andrade1}, having as the main goal to formally demonstrate 
their equivalence, relies on general and very technical arguments, making  
difficult to fully appreciate the conceptual features associated to 
the QWs reciprocities.
Thus, knowing that CQWs and SQWs are always equivalent, we can focus on 
representative cases, discussing in details the interplay between 
the discrete time formulations.
Hence, here we illustrate that different QWs are in fact closely tied 
and a specific model choice might be much more a matter of practicality 
than of unfeasibility of the corresponding sibling construction. 

The work is organized as the following. 
From a parallel with usual classical random walks, in Sec. 2 we discuss 
for the 1D case the main characteristics of CQWs and SQWs, also addressing
their mapping.
We do so assuming the most general situation of complete arbitrary
position dependent transition probabilities at the 1D lattice sites.
In Sec. 3, the same type of analysis is carried out for two 2D lattices, 
the square and the honeycomb.
The models are formulated in details and simulations, exemplifying their 
distinctions for the time evolution along the graphs structures, are 
presented.
Importantly, we show how the spatial probabilities distributions for one
model can be obtained from the other through proper projections.
Finally, results for different coin (and the equivalent scattering) 
matrices operators are compared.
It includes the more common Hadamard, Grove and Discrete Fourier Transform, 
as well as the not so usual (but interesting) Discrete Hartley Transform
and few others.
A short conclusion is drawn in Sec. 4.

\section{Discrete quantum walks on the line: two different views}

To better understand the main ideas underlying the construction of the 
existing discrete QWs models, we first recall an interesting way to 
view the classical case on the line (1D).

Consider a classical walker, starting in $x=0$ at $t=0$, that takes steps 
of fixed length $L$ and moves with constant speed $v$.
Its simple dynamics is described as the following.
Each time the walker reaches a position
$x(t = n \tau) = x_n = \pm j L$ ($j=0,1,\ldots$) for
$n = 0,1,\ldots$ and $\tau = L/v$, it randomly (and instantaneously) 
chooses a new direction to go, either to the right 
($\sigma = +1$), with probability $p$, or to the left ($\sigma = -1$), 
with probability $1-p$.
So, this 1D continuous system can be characterized by two processes taking 
place in an ``effective lattice'':
(i) one purely stochastic (choosing directions), occurring at 
the ``lattice sites'' $\pm j L$; and
(ii) other purely deterministic (ballistic motion, with
$x(t) = x_{n-1} + \sigma v (t - t_{n-1})$ for $(n-1) \tau \leq t \leq n \tau$), 
occurring along the ``lattice bonds''.
Such view is schematically represented in Fig. 1.

The construction of a discrete time QW closely follows the previous picture.
Essentially, there are two possible implementations, based on what we 
assume as the primary process, either (i) or (ii) above, to describe 
the quantum states.
Indeed, in the first case (associated to CQWs) the quantum states describe 
the system at the lattice site positions: classically, the locations where 
it is made a probabilistic choice about the next step direction.
In the second case (associated to SQWs) the quantum states are defined on 
the bonds: classically, corresponding to the deterministic locomotion along 
the lattice.
This is an important distinction since each model has a different state 
representation with a different possible interpretation.

\begin{figure}
\centerline{\psfig{figure=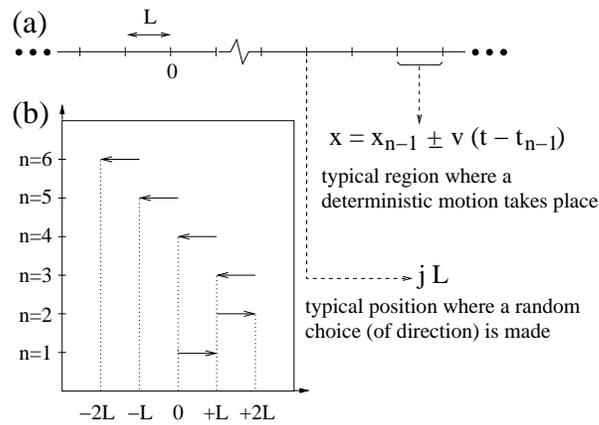,width=8cm}}
\caption{(a) A classical random walk faced as alternate stochastic 
and deterministic processes along a 1D lattice.
(b) Example of a possible path for $t = 6 \tau$, whose probability is 
$p^2 (1-p)^4$. 
Here, $p$ [$1-p$] is the probability to go to right [left].}
\label{fig1}
\end{figure}

\begin{figure}
\centerline{\psfig{figure=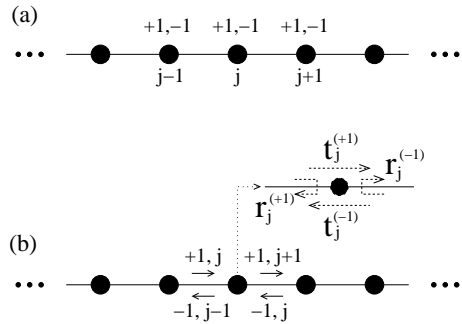,width=6cm}}
\caption{For QWs in 1D, the associated ``Hilbert lattice'', 
but which is not necessarily a spatial structure once the states do 
not need to be position eigenvectors.
The states are defined (a) on the sites, coined, and 
(b) along the bonds, scattering, formulations.
For SQWs, details in (b) show the scattering quantum amplitudes defined 
on each site.}
\label{fig2}
\end{figure}

One does not need to relate any  ``lattice'' structure to QWs.
Nevertheless, it is instructive to go on with such analogy and to view 
quantum walks in 1D as a dynamics defined on a ``Hilbert lattice''
\cite{andrade-2011}, depicted in Fig. 2.
In this way, we can associate the classical step length $L$ to the 
characteristic unit distance $\Delta{j} = 1$ between two consecutive 
``sites'' of the Hilbert lattice.
Moreover, in both cases the discrete time evolution is given by a 
single step unitary operator $U$, acting on states belonging to
this Hilbert (lattice) space, so that
$| \Psi(n+1) \rangle = U | \Psi(n) \rangle$.

\subsection{The coined quantum walk model}

Quantum states are defined on the sites $j$, Fig. 2 (a).
Then, we consider the basis states $\{|j\rangle\}$, which span the Hilbert 
subspace $\mathcal{H}_p$.
Moreover, to any $j$ we associate an inner two-level Hilbert subspace 
$\mathcal{H}_c$, whose ``coin'' states are $| \pm \rangle$.
They give the correct unitary ``stochastic'' character for the 
problem~\cite{jsp.1996.85.551, ambainis-2003}.
Also, their role is somehow similar to that of choosing directions in 
the classical case: once $|j \rangle$ represents only spatial ``location'' 
(sites), then ``direction'' must be played by the auxiliary coin states.

The appropriate orthonormal basis to describe the entire space 
$\mathcal{H} = \mathcal{H}_p \otimes \mathcal{H}_c$ 
($L^2(\mathbb{Z}) \otimes L^2(\mathbb{Z}_2)$) is
$\{|j\rangle \otimes |\sigma\rangle\}$, with $\sigma = \pm$ and 
$j = 0, \pm 1, \pm 2, \ldots$.
Defining a shift operator as
$S | j \rangle = | j + 1 \rangle$ (and $S^{\dagger} 
| j \rangle = | j - 1 \rangle$), we write
\begin{equation}
U_{c} = \left( 
         S \otimes |+ \rangle\langle + | +
	 S^{\dagger} \otimes |- \rangle\langle - |  
    \right)
	 \left(\sum_j |j \rangle \langle j| \otimes C^{(j)}
    \right).
\label{u-coined}
\end{equation}
$U_{c}$ is unitary if the coin-flip operators $C^{(j)}$'s are also unitary.
There are many possible choices for the $C$'s \cite{prl.2004.93.190503}, 
a common one being the Hadamard's, for which $\forall$ $j$,
$C^{(j)} = H_2$ with $H_2 |\sigma \rangle =
(|-\sigma \rangle - \sigma |\sigma \rangle)/\sqrt{2}$.
However, the most general case is to consider the matrix 
elements 
$\langle \sigma'' | C^{(j)} | \sigma' \rangle
= c^{(j)}_{\sigma'' \, \sigma'}$ and to assure unitarity
just by imposing
\begin{eqnarray}
& &
|c^{(j)}_{+ \, +}|^2 + |c^{(j)}_{- \, +}|^2 = 
|c^{(j)}_{- \, -}|^2 + |c^{(j)}_{+ \, -}|^2 = 1, \ \
|c^{(j)}_{+ \, -}|^2 = |c^{(j)}_{- \, +}|^2, 
\nonumber \\
& & 
c^{(j)}_{+ \, +} \, [c_{- \, +}^{(j)}]^{*} + 
c^{(j)}_{+ \, -} \, [c^{(j)}_{- \, -}]^{*} = 0.
\label{coin-condition}
\end{eqnarray}
Thus, we have 
\begin{equation}
U_{c} | j \rangle \otimes | \sigma \rangle = 
c^{(j)}_{\sigma \sigma} \;
| j + \sigma \rangle  \otimes | \sigma \rangle+  
c^{(j)}_{-\sigma \, \sigma} \;
| j - \sigma \rangle \otimes | -\sigma \rangle.
\label{evolution-coin}
\end{equation}

One of the most important applications of CQWs is in quantum computation.
So, it is usual to define the above matrix elements in a way easy to 
associate to quantum gate operators.
It explains, e.g., the frequent choice in the literature of the Hadamard's 
coin (fundamental to manipulate q-bits).
This practice, however, leads to constructions of coin matrices in a 
format not common among physicists, but popular in mathematics and 
computer science (as nicely explained and exemplified in Ref. 
\cite{nielsen}, e.g., p. xxiii).
Indeed, the states $| \pm  \rangle$ are written as the column matrices
\begin{equation}
| - \rangle = \left(
    \begin{array}{c}
      1
\\
      0
    \end{array}
\right), \qquad 
| + \rangle = \left(
    \begin{array}{c}
      0
\\
      1
    \end{array}
\right),
\end{equation}
opposite to the usual spin-up and spin-down convention.
Following this representation, the most general form for the coin 
operator, obeying to Eq. (\ref{coin-condition}) and having the Hadamard 
as a particular case is (see also
\cite{bach-2004,carneiro-2005,pra.2008.77.032326})
\begin{equation}
C^{(j)} = \exp[i \gamma_j] 
\left(
    \begin{array}{cc}
      \exp[i\xi_j]    \cos[\theta_j]  &  \exp[i\zeta_j] \sin[\theta_j]   
\\
      \exp[-i\zeta_j] \sin[\theta_j]  & -\exp[-i\xi_j]  \cos[\theta_j]
    \end{array}
\right).
\label{eq:U2_coin}
\end{equation}
with $0 \leq \gamma_j, \zeta_j, \xi_j, \theta_j < 2 \pi$.
Then, Eq. (\ref{evolution-coin}) reads
%
%
\begin{eqnarray}
U_{c} | j \rangle \otimes | \sigma \rangle &=& 
- \sigma \exp[i (\gamma_j - \sigma \xi_j)] \cos[\theta_j] \; 
| j + \sigma \rangle \otimes | +\sigma \rangle
\nonumber \\
& & +  \exp[+ i (\gamma_j + \sigma \zeta_j)] \sin[\theta_j] \; 
| j - \sigma \rangle \otimes | -\sigma \rangle.
\label{evolution-coin-particular}
\end{eqnarray}

\subsection{The scattering quantum walk model}

For our purposes, it is easier to formulate the scattering model in a 
slightly different (but completely equivalent) way than usually 
done in the literature~\cite{hillery-2007}.
Along each bond, joining two consecutive sites, say $j-1$ and $j$ 
(see Fig. 2 (b)), we assume two possible states $|+1,j \rangle$ and 
$|-1,j-1 \rangle$. 
Hence, contrasting to the coined model, now the quantum number 
$\sigma = \pm 1$ is associated to a ``direction'' along a bond in the 
Hilbert lattice.
The full Hilbert space $\mathcal{H}$ is no longer a direct product of 
two subspaces.
Each basis element $|\sigma, j \rangle$ belongs to 
$L^2(\mathbb{Z} \times \mathbb{Z}_2)$ and satisfies to
$\langle j', \sigma'|\sigma, j \rangle = 
\delta_{j'  j} \, \delta_{\sigma' \sigma}$.

Defining the operators $T$ and $R$ by
\begin{equation}
T |\sigma,j \rangle =
t^{(j)}_{\sigma} \, |\sigma, j + \sigma \rangle,
\
R |\sigma,j \rangle = 
r^{(j)}_{\sigma} \, |-\sigma, j - \sigma \rangle,
\end{equation}
and
\begin{equation}
T^{\dagger} |\sigma,j \rangle =
{t^{(j-\sigma)}_{\sigma}}^{*} \, |\sigma,j -\sigma \rangle,
\
R^{\dagger} |\sigma,j \rangle =
{r^{(j-\sigma)}_{-\sigma}}^{*} \, |-\sigma, j - \sigma \rangle,
\label{r-t-dagger-1d}
\end{equation}
the one step time evolution is simply
\begin{equation}
U_{s} = T + R,
\label{evolution-qrw}
\end{equation}
so that
\begin{equation}
U_{s} |\sigma, j \rangle = 
t^{(j)}_\sigma \, |\sigma, j + \sigma \rangle +
r^{(j)}_\sigma \, |-\sigma, j - \sigma \rangle.
\label{evolution-scattering}
\end{equation}
The unitarity of $U_{s}$ implies \cite{schmidt1}
\begin{equation}
|t^{(j)}_{\sigma}|^2 + |r^{(j)}_{\sigma}|^2 = 1,
\ \ \ \ 
|r^{(j)}_{\sigma}|^2 = |r^{(j)}_{-\sigma}|^2, \ \ \ \
r^{(j)}_{-\sigma} \, {t^{(j)}_{\sigma}}^{*} + 
{r^{(j)}_{\sigma}}^{*} \, t^{(j)}_{-\sigma} = 0,
\label{condition-scattering}
\end{equation}
which are exactly the relations satisfied by the reflection and 
transmission amplitudes in a quantum scattering problem in 1D 
\cite{schmidt1, chadan}, resulting from the unitarity of the $S$ 
scattering matrix. 

Equation (\ref{condition-scattering}) automatically holds if for any $j$ 
(with $0 \leq \rho_j \leq 1$ and 
$0 \leq \lambda_j, \phi_j, \varphi_j < 2 \pi$)
\begin{eqnarray}
t^{(j)}_{\sigma} &=& \exp[i \lambda_j] \left( 
\sqrt{1-\rho_j} \, \exp[i \sigma \phi_j] \right), 
\nonumber \\
r^{(j)}_{\sigma} &=& \exp[i \lambda_j] \left(
\sigma \, \sqrt{\rho_j} \, \exp[i \sigma \varphi_j]\right). 
\label{eigenvalues}
\end{eqnarray}
If $\lambda_j = \lambda$ for all $j$, without loss of generality we can
set $\lambda = 0$.
We should mention that the expressions in Eq. (\ref{eigenvalues}) are not
the only possibility \cite{andrade-2011}.
There is a certain arbitrariness in signals convention.
The present choice, however, has a direct physical motivation.
For time-reverse invariant systems $t^{(j)}_{+} = t^{(j)}_{-}$ \cite{chadan}.
Then, if we also assume real amplitudes, the signals for 
the $r$'s must be opposite, exactly the case in Eq. 
(\ref{eigenvalues}).
Furthermore, for $j$-independent scattering coefficients, the above (with 
$\phi_j = \varphi_j = 0$) are the relations adopted in the original work 
introducing SQW models~\cite{hillery-2003}.

\subsection{Obtaining the probabilities}

At this point we should emphasize that the above constructions are more 
general than simply to quantize the dynamics of a common classical random 
walk.
Here, by common we mean that each time a new direction 
needs to be chosen, we use the same probabilities ($p$ and $1-p$) to decide 
between right and left.
Quantum mechanically, to allow the parameters to depend on $j$ (cf., Eqs. 
(\ref{eq:U2_coin}) and (\ref{eigenvalues})) implies that we explicit assume 
position dependent probability amplitudes.
Obviously, by making such parameters $j$-independent, we recover the 
relation with the usual classical case. 

Now, suppose we shall determine which is the probability $P^{(j)}(n)$ to 
be in the ``position'' state $j$ (which means a site (bond) state in the 
coin (scattering) model) at time $n$, regardless the value of the coin 
(direction) quantum number $\sigma$.
We define then
\begin{equation}
\mathcal{P}^{(j)}_{c} =
| j \rangle \langle j | \otimes \sum_{\sigma}
| \sigma \rangle \langle \sigma |, \ \
\mathcal{P}^{(j)}_{s} = \sum_{\sigma}
| \sigma, \, j + \frac{\sigma - 1}{2} \rangle
\langle j + \frac{\sigma - 1}{2}, \, \sigma |,
\label{projectors-1d}
\end{equation}
respectively, the coin and scattering position projector operators.
The sought probability is the expected value 
\begin{equation}
P^{(j)}(n) = \langle \Psi(n) | \mathcal{P}^{(j)} | \Psi(n) \rangle,
\label{projection-1d}
\end{equation}
for ${\mathcal P}^{(j)}$ one of the expressions in Eq. (\ref{projectors-1d}).

Projection is thus an essential ingredient in defining a QW model, which 
differs from classical walk systems by the typical interference effects due 
to the Eq. (\ref{projection-1d}) (see, e.g., Ref. 
\cite{andrade-interferencia}).

\subsection{The unitary equivalence of the two 1D quantum walk models}

Finally, in the present 1D topology the unitary equivalence between CQWs and 
SQWs is straightforwardly established.

First, note the one-to-one correspondence between their full
Hilbert spaces, $L^2(\mathbb{Z}) \otimes L^2(\mathbb{Z}_2) \equiv
L^2(\mathbb{Z} \times \mathbb{Z}_2)$.
So, it follows directly the existence of an isomorphic unitary operator
\cite{hillery-2003} $E: {\mathcal H} \rightarrow {\mathcal H}$, given by 
\begin{equation}
E \, |\sigma, j \rangle = 
| j \rangle  \otimes | \sigma \rangle, \qquad
E^{\dagger} \, | j \rangle  \otimes | \sigma \rangle 
= |\sigma, j \rangle.
\label{states-mapping-1d}
\end{equation}

Second,
(i) inspecting Eqs. (\ref{evolution-coin}) and (\ref{evolution-scattering});
(ii) considering the most general conditions for the models quantum 
coefficients, represented by Eqs. (\ref{coin-condition}) and 
(\ref{condition-scattering}) (observe that Eqs. (\ref{eq:U2_coin}) and 
(\ref{eigenvalues}), more usual in the literature, already obey particular 
conventions);
and (iii) taking into account Eq. (\ref{states-mapping-1d}); one 
finds that the following coefficients play complete similar roles
\begin{equation}
c^{(j)}_{\sigma \, \sigma} \leftrightarrow t^{(j)}_{\sigma}
\ \ \ \mbox{and} \ \ \
c^{(j)}_{-\sigma \, \sigma} \leftrightarrow r^{(j)}_{\sigma}.
\label{correspondence-1d}
\end{equation}
Hence, setting $c_{\sigma \, \sigma}^{(j)} = t_{\sigma}^{(j)}$ and 
$c_{-\sigma \, \sigma}^{(j)} = r_{\sigma}^{(j)}$ for all $j$, we have in both models 
exactly the same probability amplitudes for their time evolutions.

Third, from Eqs. (\ref{evolution-coin}), (\ref{evolution-scattering}), 
(\ref{states-mapping-1d}) and assuming the above equalities, the resulting 
dynamics are unitary equivalent once
\begin{equation}
U_{s} = E^{\dagger} \, U_{c} \, E.
\label{unitary-equivalence-1d}
\end{equation}

As it is known \cite{hillery-2003}, even when different QW models are 
unitary equivalent, the associated probabilities -- obtained through 
direct projections -- can be distinct.
This is so because we are choosing contrasting {\em physical} 
representations, sites (coined) and bonds (scattering), to describe the 
problem.
The specific states which characterize one of these ``spatial'' 
configurations are not akin to the states for the other.
For instance, the two states corresponding to the site $j$ in CQWs,
$| j \rangle \otimes |\sigma = \pm \rangle$, are mapped to states at 
different bonds in SQWs.
Therefore, the probability to be in a unique site is not equal to the 
probability to be in a unique bond (cf., Eqs. 
(\ref{projectors-1d})-(\ref{projection-1d})).
Mathematically, it is related to the fact that considering Eqs. 
(\ref{projectors-1d}) and (\ref{states-mapping-1d}), we get
$\mathcal{P}_s^{(j)} \neq E^{\dagger} \, \mathcal{P}_{c}^{(j)} \, E$ and
$\mathcal{P}_c^{(j)} \neq E \, \mathcal{P}_{s}^{(j)} \, E^{\dagger}$ 
(inequalities holding true in any topology given the results in 
\cite{andrade1}).
Thus, for $|\Psi\rangle_s$ and $|\Psi\rangle_c = E \, |\Psi\rangle_s$,
we have from Eq. (\ref{projection-1d}) that generally 
$P^{(j)}_s \neq P^{(j)}_c$.

Nevertheless, due to the QWs unitary equivalence we can recover the 
probabilities from each other version by constructing proper cross 
projector operators.
Indeed, defining
\begin{equation}
\mathcal{P}^{(j)}_{s}\big|_{c} = 
E^{\dagger} \, \mathcal{P}^{(j)}_{c} \, E, 
\qquad
\mathcal{P}^{(j)}_{c}\big|_{s} =
E \, \mathcal{P}^{(j)}_{s} \, {E}^{\dagger},
\label{projections-mapping-1d}
\end{equation}
one readily obtains the probabilities for the coined (scattering) model
by applying $\mathcal{P}^{(j)}_{s}\big|_{c}$ ($\mathcal{P}^{(j)}_{c}\big|_{s}$)
to the state $|\Psi(n)\rangle_s$ ($|\Psi(n)\rangle_c$), evolved according
to the scattering (coined) formulation.

The numerical examples for 2D lattices in Section 3 will clearly illustrate 
all the above observations.

\section{More general topologies: two illustrative examples}

As already mentioned, it has been proven in Ref. \cite{andrade1} that CQWs 
and SQWs are always related by an unitary transformation.
In the previous Section, profiting from its relative simplicity (although
considering the most general situation, hence extending the results of 
Ref. \cite{hillery-2003}), we have explicit illustrated so for the 1D case.
 
However, the benefits in being able to map CQWs and SQWs become really 
evident when one considers more complex topologies.
Thus, next we present a detailed analysis of the correspondence between 
CQWs and SQWs in two particular, but very instructive, 2D examples,
square and honeycomb lattices.

\subsection{Quantum walks on a square lattice}

The square lattice, represented in Fig. 3, is the natural 2D extension 
of the 1D topology. 
In the following we describe in such a case the coined and scattering QW 
formulations, as well as their unitary equivalence.

\begin{figure}
\centerline{\psfig{figure=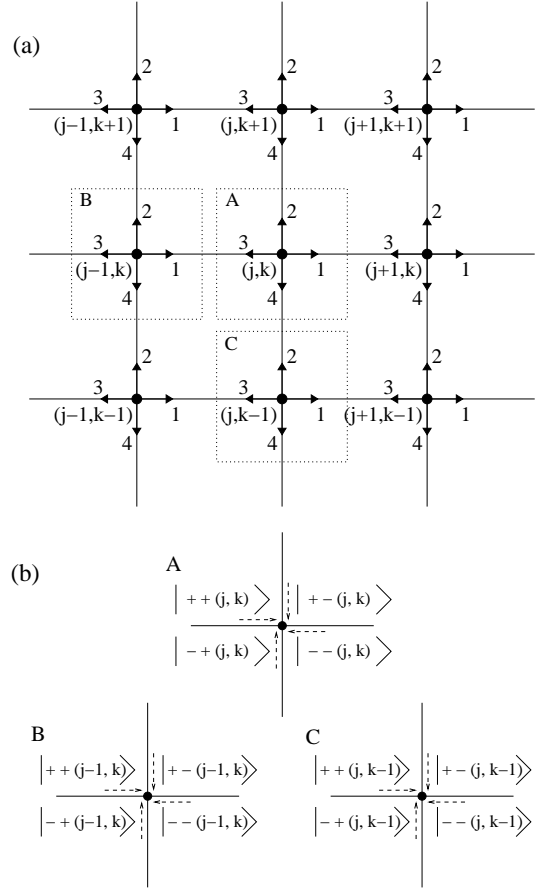,width=7cm}}
\caption{A QW on a square lattice.
(a) At each site $(j, k)$, the arrows schematically represent the four 
inner states, $\sigma = 1, 2, 3, 4$, of the coined formulation.
(b) For the scattering formulation, it is shown the four ``incoming'' 
states propagating to the lattice sites marked in (a).
The two possible states defined on the bond common to the regions A and B 
(A and C) are
$| + +, (j, k) \rangle$ and $| - -, (j-1, k) \rangle$
($| - +, (j, k) \rangle$ and $| + -, (j, k-1) \rangle$).}
\label{fig3}
\end{figure}

\begin{figure}[]
\centerline{\psfig{figure=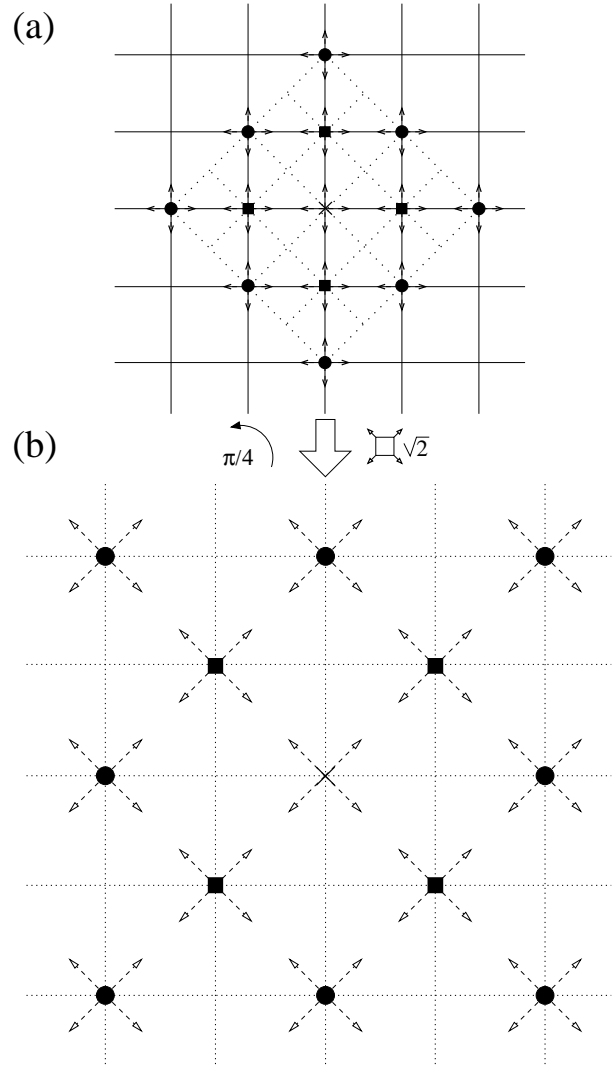,width=8cm}}
\caption{Schematics of the one step time evolution for the coined model 
in a square lattice in the natural (a) and the diagonal (b) cases. 
The system, initially at the state represented by the site $\times$, in 
a first step can reach the states represented by the sites 
$\scriptscriptstyle{\blacksquare}$.
In the second step, the states which can be visited are those 
represented by the sites ${\bullet}$ and $\times$.
The diagonal version (dotted lattices) can be thought as the natural one 
rotated by $\pi/4$ and rescaled by a factor $\sqrt{2}$.}
\label{fig4}
\end{figure}

\subsubsection{The coined formulation} \

In Fig. 3 (a) we schematically represent the coined Hilbert lattice, 
whose basis states are
$\{| j \rangle \otimes | k \rangle \otimes | \sigma \rangle\}$, with
$\sigma = 1, 2, 3, 4$ and $j, k = 0, \pm 1, \pm 2, \ldots$.
For notation simplicity, we do not write the ``spatial'' states in the 
form $| j \rangle_x \otimes | k \rangle_y$ (and operators as 
$A_x \otimes B_y$) to distinguish between the two distinct 1D dimensions.
Also, we assume the natural ordering convention
$(A \otimes B \otimes C) \, 
| \alpha \rangle \otimes | \beta \rangle \otimes | \gamma \rangle 
= A | \alpha \rangle \otimes 
  B | \beta \rangle \otimes 
  C | \gamma \rangle$.

For the inner coin states $| \sigma \rangle$, written as
\begin{equation}
| 1 \rangle = 
\left(
    \begin{array}{c}
  0 \\
  1 \\
  0 \\
  0
\end{array}
\right),  
| 3 \rangle = 
\left(
    \begin{array}{c}
  1 \\
  0 \\
  0 \\
  0
\end{array}
\right),
| 2 \rangle = 
\left(
    \begin{array}{c}
  0 \\
  0 \\
  0 \\
  1 
\end{array} 
\right),
| 4 \rangle = 
\left(
    \begin{array}{c}
  0 \\
  0 \\
  1 \\
  0
\end{array}
\right),
\end{equation}
the coin operator matrix (at each $(j,k)$) reads 
\begin{equation}
C^{(j, k)} =  
\left(
    \begin{array}{cccc}
      c^{(j, k)}_{3 \, 3}  &  c^{(j, k)}_{3 \, 1} & 
      c^{(j, k)}_{3 \, 4}  &  c^{(j, k)}_{3 \, 2} \\
      c^{(j, k)}_{1 \, 3}  &  c^{(j, k)}_{1 \, 1} & 
      c^{(j, k)}_{1 \, 4}  &  c^{(j, k)}_{1 \, 2} \\
      c^{(j, k)}_{4 \, 3}  &  c^{(j, k)}_{4 \, 1} & 
      c^{(j, k)}_{4 \, 4}  &  c^{(j, k)}_{4 \, 2} \\
      c^{(j, k)}_{2 \, 3}  &  c^{(j, k)}_{2 \, 1} & 
      c^{(j, k)}_{2 \, 4}  &  c^{(j, k)}_{2 \, 2} 
    \end{array}
\right).
\label{coin-four}
\end{equation}
The unitarity of $C$, directly extending the relations in Eq. 
(\ref{coin-condition}), leads to 
\begin{equation}
\sum_{l} c^{(j, k)}_{i \, l} {c^{(j, k)}_{m \, l}}^{*} =
\sum_{l} c^{(j, k)}_{l \, i} {c^{(j, k)}_{l \, m}}^{*} = 
\delta_{i \, m}.
\label{coin-condition-square}
\end{equation}

As the one step time evolution, there are two common possibilities.
The first is the so called natural choice, since single steps 
displacements follow the natural topology of the lattice \cite{mackay},
i.e., from a given site representing a state, it can go either to right, 
left, up or down \cite{tregenna1,carneiro} (Fig. 4 (a)).
The second is the diagonal version \cite{diagonal}.
In this case, single steps are those in which ``moves'' take place along 
the diagonals of the Hilbert lattice (Fig. 4 (b)).

Let us define (for $S$ as in Sec. 2-1)
\begin{equation}
N = 
S           \otimes {\mathbb I} \otimes  |1 \rangle\langle 1 | +
S^{\dagger}   \otimes {\mathbb I} \otimes  |3 \rangle\langle 3 | +
{\mathbb I} \otimes S           \otimes  |2 \rangle\langle 2 | +
{\mathbb I} \otimes S^{\dagger}   \otimes  |4 \rangle\langle 4 |,
\end{equation}
with ${\mathbb I}$ the identity operator.
The time evolution for the natural version, an extension of Eq. 
(\ref{u-coined}), yields
\begin{equation}
U_{c} = N \times
    \Big( \sum_{j, k} | j \rangle \langle j | \otimes 
| k \rangle \langle k | \otimes C^{(j, k)} \Big).
\label{u-coined-2d}
\end{equation}

For the diagonal version, we first define the operator
\begin{equation}
D = 
{\mathbb I}  \otimes S           \otimes   |1 \rangle\langle 1 | +
{\mathbb I}  \otimes S^{\dagger}   \otimes   |3 \rangle\langle 3 | +
S^{\dagger}    \otimes {\mathbb I} \otimes   |2 \rangle\langle 2 | +
S            \otimes {\mathbb I} \otimes   |4 \rangle\langle 4 |.
\label{d-rotation}
\end{equation}
So, the time evolution is given by
\begin{equation}
U_{c-diagonal} = D \, U_{c}.
\label{u-coin-diagonal}
\end{equation}
Equation (\ref{u-coin-diagonal}), with $D$ as in Eq. (\ref{d-rotation}), has 
a very simple interpretation (if we maintain that Eq. (\ref{u-coin-diagonal}) 
represents a single step).
Indeed, the diagonal case is essentially the natural one rotated by $\pi/4$ 
(anti-clockwise) in the Hilbert lattice.
Moreover, quantities associated to any norm (length scale) defined on the 
lattice, e.g., diffusiveness, should be re-scaled by a factor $\sqrt{2}$, 
which is just the ratio between the diagonal and side of the unitary cell, 
Fig. 4.
In all our further analysis, we will consider only the natural version, 
observing that the diagonal would easily follow from Eq. 
(\ref{u-coin-diagonal}).

Finally, for the natural coined model the projector operator is
\begin{equation}
{\mathcal P}^{(j, k)}_{c} =
| j \rangle \langle j | \otimes | k \rangle \langle k | \otimes 
\sum_{\sigma=1}^{4}
| \sigma \rangle \langle \sigma |.
\label{projection-coin-four}
\end{equation}

\subsubsection{The scattering model} \

For $\sigma_x, \sigma_y = \pm 1$ ($\pm$ for short) and 
$j, k = 0, \pm 1, \pm 2, ...$, the scattering model basis states are 
$\{ | \sigma_x \, \sigma_y, (j, k) \rangle \}$, Fig. 3 (b). 
The time evolution is again $U_{s} = T + R$, with 
\begin{eqnarray}
\hspace{-2cm} T |\sigma_x \, \sigma_y, (j, k) \rangle 
&=& \sum_{\{\alpha, \beta\} \neq \{-\sigma_x, -\sigma_y \}}
t_{\alpha \,  \beta , \, \sigma_x \, \sigma_y}^{(j,k)}  \,  
| \alpha \, \beta,
(j + \frac{\alpha |\alpha + \beta|}{2}, 
 k + \frac{\beta |\alpha - \beta|}{2}) \rangle,
\nonumber \\ 
\hspace{-2cm} R |\sigma_x \, \sigma_y, (j, k) \rangle 
&=&
r_{-\sigma_x \,  -\sigma_y , \, \sigma_x \, \sigma_y}^{(j,k)}  \,
|-\sigma_x \, -\sigma_y, 
(j - \frac{\sigma_x |\sigma_x + \sigma_y|}{2}, 
 k - \frac{\sigma_y |\sigma_x - \sigma_y|}{2}) \rangle.
\nonumber \\
\label{r-t-2d}
\end{eqnarray}
The corresponding actions of $T^{\dagger}$ and $R^{\dagger}$ are a 
straightforward generalization of Eq. (\ref{r-t-dagger-1d}), in a 
format similar to Eq. (\ref{r-t-2d}). 
Defining
\begin{equation}
\Gamma^{(j, k)} =  
\left(
    \begin{array}{cccc}
      t^{(j, k)}_{- - , \, - -}  &  r^{(j, k)}_{- - , \, + +} & 
      t^{(j, k)}_{- - , \, + -}  &  t^{(j, k)}_{- - , \, - +} \\
      r^{(j, k)}_{+ + , \, - -}  &  t^{(j, k)}_{+ + , \, + +} & 
      t^{(j, k)}_{+ + , \, + -}  &  t^{(j, k)}_{+ + , \, - +} \\
      t^{(j, k)}_{+ - , \, - -}  &  t^{(j, k)}_{+ - , \, + +} & 
      t^{(j, k)}_{+ - , \, + -}  &  r^{(j, k)}_{+ - , \, - +} \\
      t^{(j, k)}_{- + , \, - -}  &  t^{(j, k)}_{- + , \, + +} & 
      r^{(j, k)}_{- + , \, + -}  &  t^{(j, k)}_{- + , \, - +} 
    \end{array}
\right), 
\label{smatrix-four}
\end{equation}
the unitarity of $U_{s}$ implies in the unitarity of the scattering matrices 
$\Gamma^{(j, k)}$ (once more a direct extension of previous results, in
this case of Eq. (\ref{condition-scattering})).
So, the elements of $\Gamma^{(j, k)}$ must satisfy to relations completely 
analog to those in Eq. (\ref{coin-condition-square}).

For the projector operators, we need to distinguish between the horizontal 
($\sigma_x \times \sigma_y = +$) and the vertical 
($\sigma_x \times \sigma_y = -$) bonds.
Hence, we define 
\begin{eqnarray}
\mathcal{P}^{(j, k; +)}_{s} &=& \sum_{\sigma_x}
| \sigma_x \, \sigma_x, 
(j + \frac{\sigma_x - 1}{2}, k) \rangle
\langle (j + \frac{\sigma_x - 1}{2}, k),
\sigma_x \, \sigma_x |, \nonumber \\
\mathcal{P}^{(j, k; -)}_{s} &=& \sum_{\sigma_y}
| -\sigma_y \, \sigma_y, 
(j, k + \frac{\sigma_y - 1}{2}) \rangle
\langle (j, k + \frac{\sigma_y - 1}{2}),
-\sigma_y \, \sigma_y |.
\label{projection-scattering-four}
\end{eqnarray}

\subsubsection{The mapping} \

To map the two models, we first make the following identification 
between the scattering directions and the inner coin quantum numbers 
($\sigma_ x \, \sigma_y \leftrightarrow \sigma$):
\begin{equation}
+ + \leftrightarrow 1, \ 
- + \leftrightarrow 2, \
- - \leftrightarrow 3, \
+ - \leftrightarrow 4, 
\label{association-four}
\end{equation}
which can be cast as 
$\sigma = (5 -  (2 + \sigma_x) \sigma_y)/2$.

Using the procedure in the last Section (or likewise, the rigorous
construction in Ref. \cite{andrade1}), i.e.,  
to associate the scattering states ``incoming'' to a certain site (see 
Fig. 3 (b)) with the coin states at that site, we can set $E$ as (taking 
into account Eq. (\ref{association-four}))
\begin{equation}
E \, |\sigma_x \, \sigma_y \, (j, k) \rangle = 
| j \rangle \otimes | k \rangle \otimes 
| \frac{5}{2} -  \frac{(2 + \sigma_x) \sigma_y}{2} \rangle.
\label{states-mapping-square}
\end{equation}
Furthermore, assuming the coefficients in Eq. (\ref{coin-four}) equal to 
those in Eq. (\ref{smatrix-four}) (which is consistent with the relation 
in Eq. (\ref{association-four})), again we find that the two models are 
unitary equivalent, since for Eqs. (\ref{u-coined-2d}), (\ref{r-t-2d}) 
and (\ref{states-mapping-square}), the relation in Eq. 
(\ref{unitary-equivalence-1d}) holds.
 
Finally, to obtain the cross projector operators, we define 
$\mathcal{P}^{(j, k)}_{s}\big|_{c}$
and
$\mathcal{P}^{(j, k)}_{c}\big|_{s}$
as in Eq. (\ref{projections-mapping-1d}), for $E$ given by Eq. 
(\ref{states-mapping-square}), $\mathcal{P}^{(j, k)}_{c}$ by 
Eq. (\ref{projection-coin-four}), and $\mathcal{P}^{(j, k)}_{s}$ 
by the appropriate $\mathcal{P}^{(j, k); \pm}_{s}$ in 
Eq. (\ref{projection-scattering-four}).

\subsubsection{Examples} \

To illustrate the above general constructions, we analyze some particular 
cases for the probability amplitudes (see, for instance, Refs. 
\cite{tregenna1,mackay,carneiro}).
For simplicity, we suppose all the coin (and therefore the scattering) 
matrices to be independent on the quantum numbers $j$ and $k$.
We choose the following four coin operators (whose corresponding
scattering matrices are written in exactly the same form): \\
\\
(a) $x$--$y$ decoupled Hadamard
\begin{equation}
C_{H_2 \oplus H_2} = \left(
\begin{array}{cc}
      H_2 & 0  \\
      0   & H_2
\end{array}
\right)
= 
\frac{1}{\sqrt{2}}
\left(
\begin{array}{cccc}
      +1 & +1 & 0  &  0 \\
      +1 & -1 & 0  &  0 \\
       0 &  0 & +1 & +1 \\
       0 &  0 & +1 & -1
\end{array}
\right); 
\label{h2-decoupled}
\end{equation}
(b) Full $4 \times 4$ Hadamard
\begin{equation}
C_{H_4} = 
\frac{1}{2}
\left(
\begin{array}{cccc}
      +1 & +1 & +1 & +1 \\
      +1 & -1 & +1 & -1 \\
      +1 & +1 & -1 & -1 \\
      +1 & -1 & -1 & +1
\end{array}
\right);
\label{h4} 
\end{equation}
(c) $4 \times 4$ Grover
\begin{equation}
C_{G_4} = 
\frac{1}{2}
\left(
\begin{array}{cccc}
      -1 & +1 & +1 & +1 \\
      +1 & -1 & +1 & +1 \\
      +1 & +1 & -1 & +1 \\
      +1 & +1 & +1 & -1
\end{array}
\right);
\label{grover}
\end{equation}
(d) $4 \times 4$ Discrete Fourier Transform (DFT)
\begin{equation}
C_{DFT_4} = 
\frac{1}{2}
\left(
\begin{array}{cccc}
      +1 & +1 & +1 & +1 \\
      +1 & +i & -1 & -i \\
      +1 & -1 & +1 & -1 \\
      +1 & -i & -1 & +i
\end{array}
\right).
\label{dft}
\end{equation}

We compute $U^n \, |\Psi(0)\rangle$, with $n=20$, for both QW models and 
the above matrices.
For each $(j,k)$ we use proper projectors to calculate the probability 
$P^{(j,k)}(n=20) = \langle \Psi(20) | \mathcal{P}^{(j,k)} | \Psi(20) \rangle$.
As $|\Psi(0)\rangle$, we take
\begin{eqnarray}
|\Psi(0)\rangle_s &=& \frac{1}{2} 
\Big[
|+ + \, (0,0)\rangle + i \, |- - \, (0,0)\rangle + 
|- + \, (0,0)\rangle + i \, |+ - \, (0,0)\rangle
\Big],
\nonumber \\
|\Psi(0)\rangle_c &=& E \, |\Psi(0)\rangle_s = \frac{1}{2} \,
| 0 \rangle \otimes | 0 \rangle \otimes
\Big[ | 1 \rangle + i \, | 3 \rangle + | 2 \rangle + i \, | 4 \rangle
\Big].
\label{initial-square}
\end{eqnarray}

\begin{figure}[]
\centerline{\psfig{figure=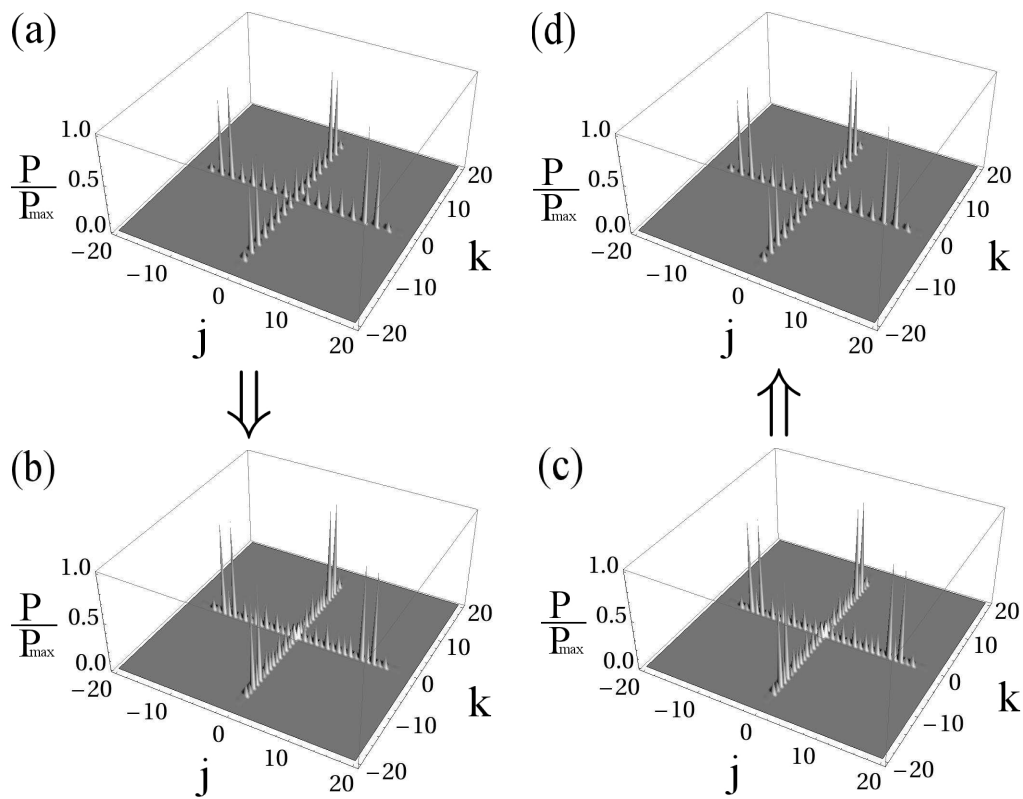,width=13.5cm}}
\caption{Quantum walks in the square lattice considering the decoupled
Hadamard transition coefficients, Eq. (\ref{h2-decoupled}).
The density plots represent the probabilities to be in the different 
states (as defined by the projector operators) after $n=20$ time steps 
for $|\Psi(0)\rangle$ given in the main text.
The results are for the: (a) coined, (b) scattering obtained from the 
coined, (c) scattering, and (d) coined obtained form the scattering,
models.
The graphs (a) and (d) and (b) and (c) are complete identical.}
\end{figure}

The $P^{(j,k)}$'s are displayed as 3D density plots.
So, in each graph the heights in the $z$--axis correspond to the 
probabilities values.
Moreover, here and for the honeycomb lattice in Sec. 3.2, each pair of 
labels $j$ and $k$ in the $x$--$y$ plane indicates the $(j,k)$ lattice site 
spatial location. 
Thus, for the coined formulation, the $P^{(j,k)}$'s are marked just over the 
sites.
For the scattering formulation, since the states are defined along the 
bonds, we mark the $P$'s exactly over the middle points of the corresponding
bonds.
Hence, we show the probability patterns of each QW version in its own 
state representation, but in a way which makes easy to qualitatively 
compare the two models.

The results for the decoupled Hadamard, full Hadamard, Grove, and DFT 
operators are presented, respectively, in Figs. (5) to (8).
The plots are organized as the following.
For the coined formulation, Figs. 5--8 (a) [(b)] show the probabilities 
obtained from the projectors in Eq. (\ref{projection-coin-four}) [the 
projectors $\mathcal{P}^{(j, k)}_{c}\big|_{s}$].
On the other hand, for the scattering formulation, Figs. 5--8 (c) [(d)] show 
the $P^{(j,k)}$'s from the projectors in Eq. (\ref{projection-scattering-four}) 
[the projectors $\mathcal{P}^{(j, k)}_{s}\big|_{c}$]. 

\begin{figure}[]
\centerline{\psfig{figure=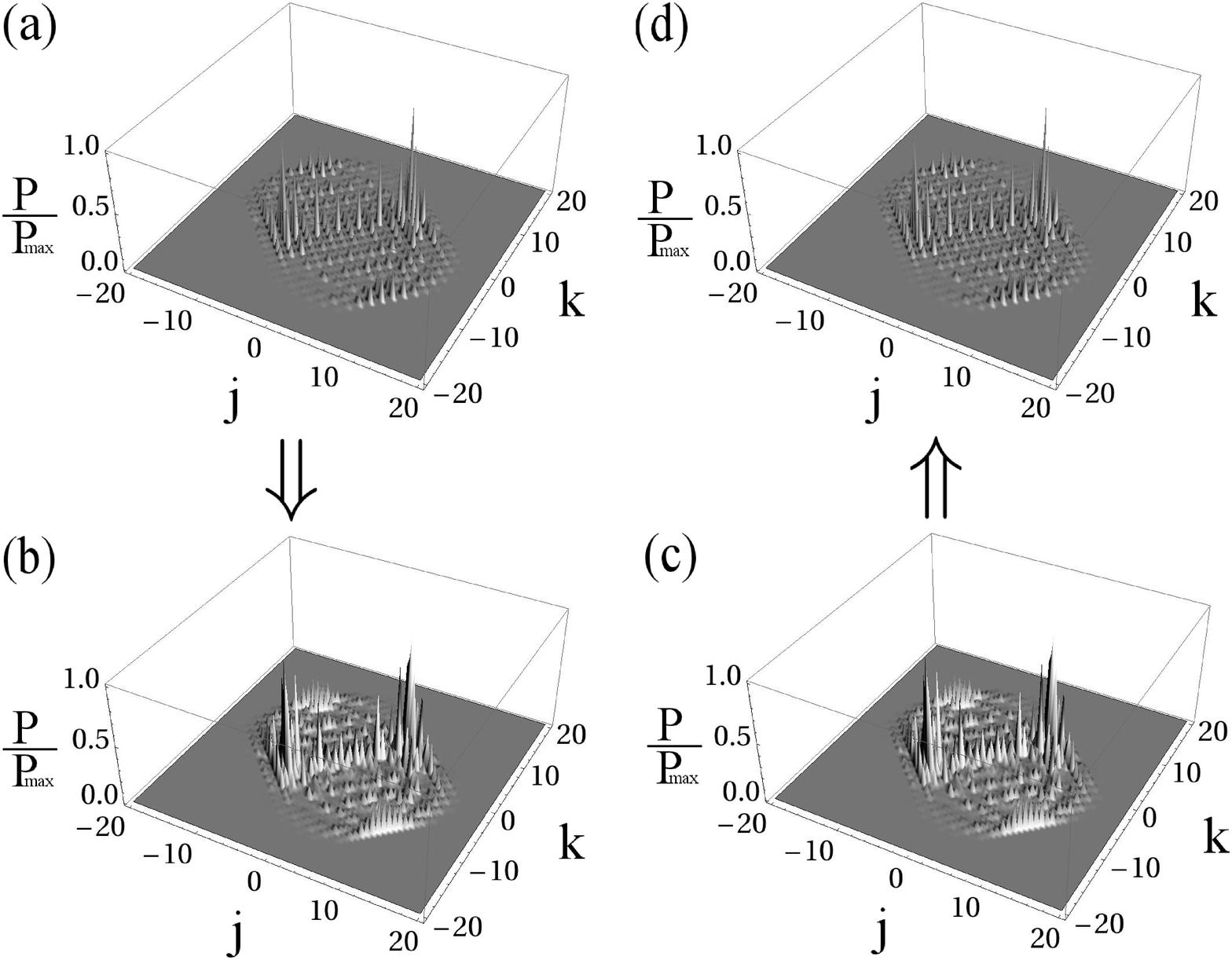,width=13.5cm}}
\caption{The same as in Fig. 5, but for the $4 \times 4$
Hadamard, Eq. (\ref{h4}).}
\end{figure}

\begin{figure}[]
\centerline{\psfig{figure=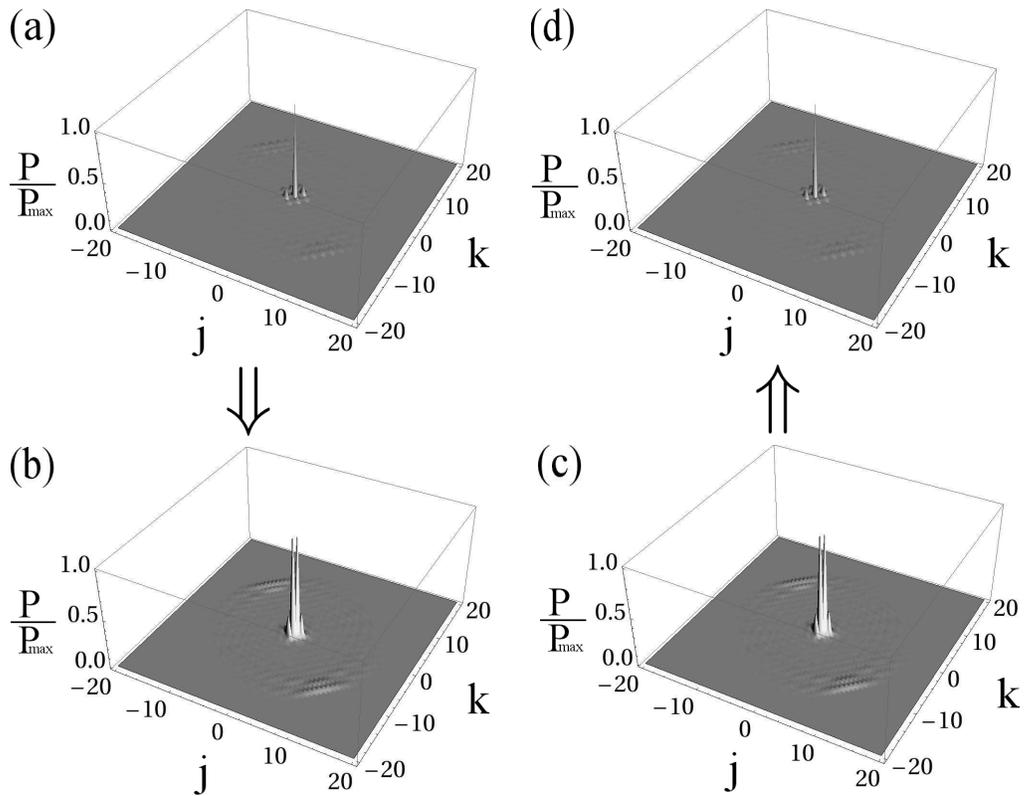,width=13.5cm}}
\caption{The same as in Fig. 5, but for the $4 \times 4$ Grover, 
Eq. (\ref{grover}).}
\end{figure}

\begin{figure}[]
\centerline{\psfig{figure=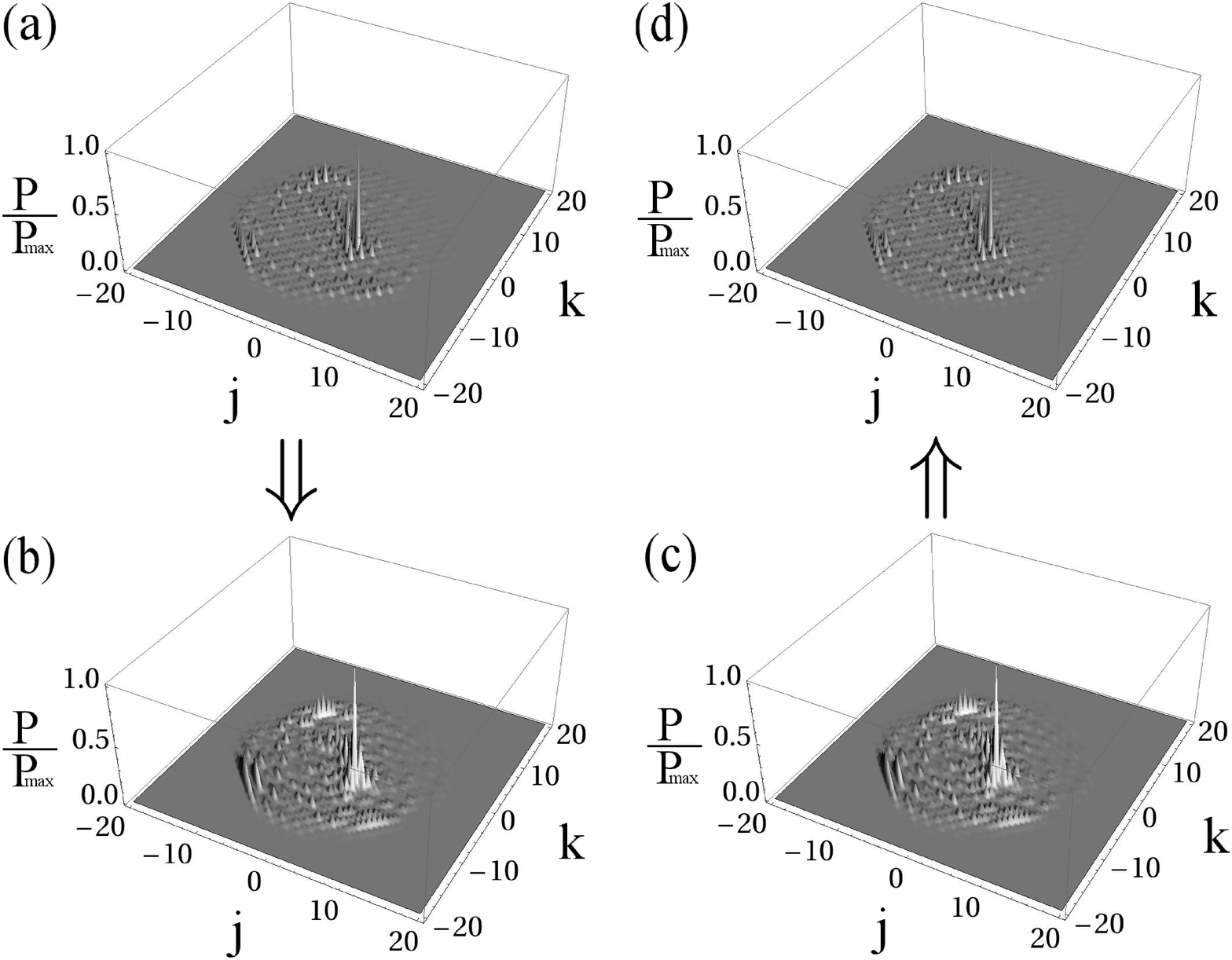,width=13.5cm}}
\caption{The same as in Fig. 5, but for the $4 \times 4$ DFT, 
Eq. (\ref{dft}).}
\end{figure}

By inspecting Figs. (5)-(8), some points become evident.
(i) Due to decoupled form of the transition amplitude matrix, Eq. 
(\ref{h2-decoupled}), in Fig. 5 we actually have two independent one 
dimensional evolutions.
Hence, each direction is a good example of the 1D results in Sec. 2.
(ii) Despite the fact the two QW versions are completely equivalent 
by unitary transformations, they lead to distinct probability patterns 
(compare (a) or (d) with (b) or (c)).
Indeed, as already emphasized (for instance, the discussion just after Eq. 
(\ref{unitary-equivalence-1d}) in Sec. 2.4), the one-to-one correspondence 
between the two formulations pinpoint a dynamical similarity. 
However, the states in each case represent different (although correlated) 
aspects of the Hilbert lattice, thus the distributions of the $P$'s do not 
need to coincide.
(iii) Moreover, such patterns display richer structures in the scattering
case.
To understand so, assume the simpler 1D lattice, for which the initial 
state $|+, 0\rangle$ ($E \, |+, 0 \rangle = | 0 \rangle \otimes |+ \rangle$)
evolves one time step.
For the SQW (CQW), we get $| \Psi(1) \rangle = 
r_+^{(0)} \, | -, -1 \rangle + t_{+}^{(0)} \, | +, +1 \rangle$
($| \Psi(1) \rangle = c_{- \, +}^{(0)} \, | -1 \rangle \otimes 
| - \rangle + c_{+ \, +}^{(0)} \, | +1 \rangle \otimes | + \rangle$).
Considering the Hilbert lattice picture, note that for the former we have 
two ``neighbor'' bond states in the expansion of $| \Psi(1) \rangle$, those
``attached'' to the site $0$.
On the other hand, for the latter the site states composing 
$| \Psi(1) \rangle$ are $j=-1$ and $j=+1$, but not $j=0$.
It illustrates a very typical situation in any topology, namely, the SQW 
dynamics tends to excite ``contiguous'' spatial (bond) basis states, 
whereas CQWs may skip some ``successive'' (site) basis states.
Thus, interference \cite{andrade-interferencia} is usually more 
recurrent in SQWs than in CQWs (due to this difference in the spreading of
$| \Psi \rangle$), explaining the behavior observed in the plots.
(iv) Nevertheless, we can recover the $P^{(j,k)}$'s from each other model by 
means of the cross projectors (e.g., SQWs in (b) from CQWs in (a) and CQWs 
in (d) from SQWs in (c)), since the information about one model is always 
encoded in the other.

We finally mention we have analyzed the diagonal CQW and its 
corresponding scattering version.
We have obtained one formulation $P$'s from the other by correctly defining 
the cross projectors.
Furthermore, as expected the probabilities plots (not shown) are exactly the 
ones here, just rotated by $\pi/4$ and rescaled by a factor $\sqrt{2}$.

\subsection{Quantum walks on a honeycomb lattice}

\begin{figure}[top]
\centerline{\psfig{figure=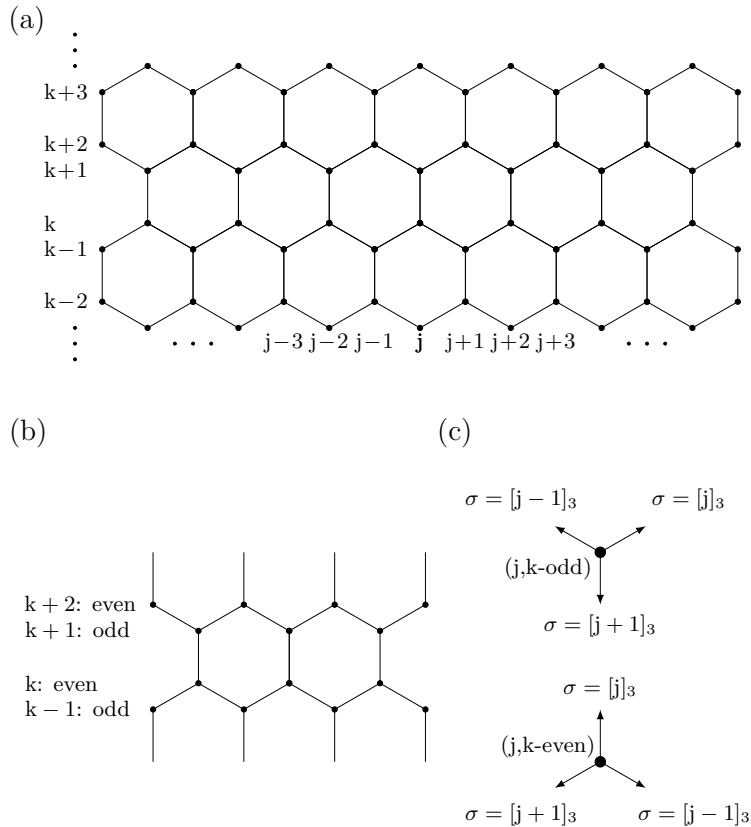,width=10cm}}
\caption{(a) The honeycomb lattice and (b) the convention for the 
sites labeling.
(c) For the coined formulation, the association between the quantum number 
$\sigma$ and directions.
Here $[x]_3$ denotes $x$ mod 3, i.e., the remainder of $x/3$.}
\end{figure}

Our last example is a QW defined on a honeycomb lattice, whose
structure is depicted in Fig. 9 (a).
It is far more involving than the previous square case mainly because 
now the coordination number, equal to 3, is odd \cite{carneiro-2005}
(for coordination 3 in 1D see, e.g., Ref. \cite{inui}).
We note that although this topology was recently investigated in the 
context of continuous time models \cite{jafarizadeh}, and few discrete 
time implementations do exist for torus-like boundary conditions \cite{abal} 
and for a similar three-state QW \cite{kollar} (but not in a truly honeycomb 
geometry), the present is the first general treatment for the problem.

The labeling of the sites in both formulations is indicated in Fig. 9 (b).
Without loss of generality, we adopt the convention: 
in the $y$-direction the quantum number $k$ is chosen such that if $k$ is 
even, then the corresponding infinite row of sites have the bonds along north 
(up), southeast and southwest. 
On the other hand, for $k$ odd, the bonds configuration at each site is 
south (down), northeast and northwest.

An interesting aspect of the honeycomb lattice, consequence of how it
imposes the relation between the quantum number $\sigma$ and directions, 
is that any sequence: 
\begin{equation}
(j_0, k_0)_\sigma 
\stackrel{U}{\rightarrow}
(j_1, k_1)_\sigma
\stackrel{U}{\rightarrow}
(j_2, k_2)_\sigma 
\, \ldots \,
\stackrel{U}{\rightarrow}
(j_{N-1}, k_{N-1})_\sigma
\stackrel{U}{\rightarrow}
(j_{N}, k_{N})_\sigma;
\label{sequence}
\end{equation}
i.e., a particular $N$ steps evolution for which the value of $\sigma$ 
remains the same, is not naturally (i.e., necessarily) a ballistic-like
trajectory.
By natural we mean those cases (like regular lattices with an even 
coordination number, e.g., our square lattice), where we always can 
associate quantum states and directions such that  
$\sigma \rightarrow \sigma$ leads to evolution along straight lines.
In such cases, the above dynamics, Eq. (\ref{sequence}), would be ballistic 
in the Hilbert lattice space regardless the specific $\sigma$.
As we are going to see for the construction adopted here, successive 
transitions $\sigma \rightarrow \sigma$ yield a return to a same site 
(in a round trip) after visiting six sites.
Hence, diffusion throughout the lattice implies transitions of 
the type $\sigma \rightarrow \sigma' \neq \sigma$.

\subsubsection{The coined model} \

For the inner coin states $| \sigma \rangle$, we have three possibilities, 
namely, $\sigma = 0, 1, 2$ (we also could call them $1, 2, 3$, but 
the use of $0$ instead of $3$ simplifies the notation).
Due to the honeycomb particular topology, some care is necessary in defining 
the dynamics in terms of the coin variable.
Indeed, states with a same $\sigma$ but at different sites does not 
always evolve to a same direction.
So, to properly associate the $\sigma$'s with the system evolution under 
$U$, we consider the following prescription (which, however, is not the 
only possible\footnote{A full classification of equivalent constructions
for the honeycomb lattice will appear elsewhere.}).
Let $[x]_N \equiv x \ \mbox{mod} \ N$, i.e., $[x]_N$ is the remainder of 
$x/N$.
Then, for a given site $(j, k)$, the quantum numbers $0, 1, 2$ are associated 
to directions in the lattice as indicated in Fig. 9 (c).
An example of the resulting configuration is displayed in Fig. 10
(in special, note from Fig. 10 that leaving from a certain site, say
$(j,k)$, and always evolving to a same $\sigma$, mandatorily will get back 
to $(j, k)$ in exactly six steps). 
The rules in Fig. 9 (c) establish in an unique and self-consistent way 
-- for the entire lattice -- how single steps $(j, k) \rightarrow (j', k')$ 
are determined by the values of $\sigma$.

\begin{figure}[top]
\centerline{\psfig{figure=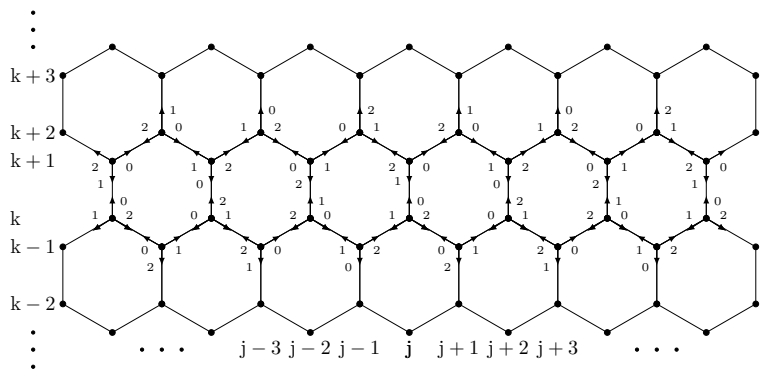,width=10cm}}
\caption{An illustration of the resulting association between coin quantum 
numbers and directions according to the rules in Fig. 9 (c).
Here $k$ is even and $[j]_3 = 0$.}
\end{figure}

The basis states are written as $|(j, k) \rangle \otimes | \sigma \rangle$,
with a column vector representation for $| \sigma \rangle$ given by
\begin{equation}
| 0 \rangle = 
\left(
    \begin{array}{c}
  1 \\
  0 \\
  0
\end{array}
\right),  
| 1 \rangle = 
\left(
    \begin{array}{c}
  0 \\
  1 \\
  0
\end{array}
\right),
| 2 \rangle = 
\left(
    \begin{array}{c}
  0 \\
  0 \\
  1 
\end{array} 
\right).
\end{equation}
So, the coin operator at any site $(j, k)$ is
\begin{equation}
C^{(j, k)} =  
\left(
    \begin{array}{ccc}
      c^{(j, k)}_{0 \, 0}  &  c^{(j, k)}_{0 \, 1} & c^{(j, k)}_{0 \, 2}  \\
      c^{(j, k)}_{1 \, 0}  &  c^{(j, k)}_{1 \, 1} & c^{(j, k)}_{1 \, 2}  \\
      c^{(j, k)}_{2 \, 0}  &  c^{(j, k)}_{2 \, 1} & c^{(j, k)}_{2 \, 2}  
    \end{array}
\right),
\label{coin-tres}
\end{equation}
which we suppose to be an unitary matrix.

Assuming the above construction, after a little lengthy but straightforward 
analyzes one finds that the one step time evolution operator for the coined 
formulation reads
\begin{equation}
U_{c} = \left( \sum_{\sigma=0}^2 S_{\sigma} \otimes
|\sigma \rangle\langle \sigma | \right) \times
    \Big( \sum_{j, k} | (j, k) \rangle \langle (j, k) | 
\otimes C^{(j, k)} \Big), 
\end{equation}
where
\begin{eqnarray}
S_{\sigma} \, |(j, k) \rangle &=& 
\left| (f(j,k;\sigma), g(j,k;\sigma)) 
\right \rangle, 
\nonumber \\
{S_{\sigma}}^{\dagger} \, |(j, k) \rangle &=& 
\left| (f(j,k;\phi_k(\sigma)), g(j,k;\phi_k(\sigma))) 
\right \rangle,
\end{eqnarray}
for (with $\mbox{sgn}[x] = x/|x|$, if $x \neq 0$, and 
$\mbox{sgn}[0] = 0$)
\begin{eqnarray}
f(j,k;\sigma) &=& j + (-1)^{\sigma- [j + [k]_2]_3}\,
\mbox{sgn}\big[ \sigma - [j + [k]_2]_3 \big], 
\nonumber \\
g(j,k;\sigma) &=& k + (-1)^k \, \big(1 -  2 \,
\mbox{sgn}\left[ |\sigma - [j + [k]_2]_3 | \right]
\big),
\nonumber \\
\phi_k(\sigma) &=& [\sigma - (-1)^k]_3.
\label{f-g}
\end{eqnarray}
By construction $S_{\sigma} \, {S_{\sigma}}^{\dagger} =
{S_{\sigma}}^{\dagger} \, S_{\sigma} = {\mathbb I}$,
leading to an unitary $U_{c}$.


Finally, the corresponding projector operator is
\begin{equation}
\mathcal{P}_{c}^{(j, k)} = 
|(j, k) \rangle \langle (j, k) | \otimes 
\sum_{\sigma=0}^{2}
| \sigma \rangle \langle \sigma |.
\label{projector-honeycomb-coin}
\end{equation}

\subsubsection{The scattering model} \

\begin{figure}[top]
\centerline{\psfig{figure=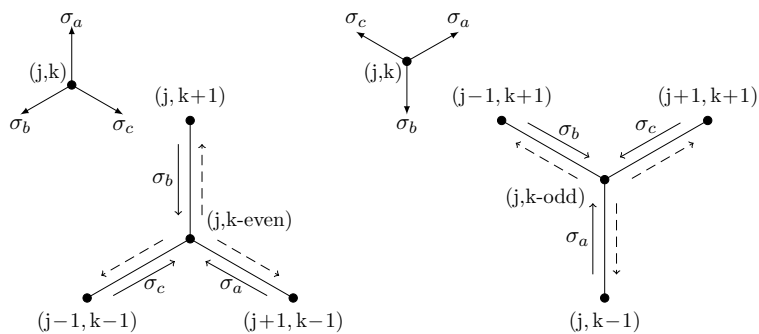,width=10cm}}
\caption{For the honeycomb lattice, the SQW basis states labeling rules.
The $\sigma$ values here ($\sigma_a$, $\sigma_b$ and $\sigma_c$) are 
defined in Eq. (\ref{sigma-convention}). 
The upper details summarize the convention in Fig. 9 (c) to label the 
CQW inner states.}
\end{figure}

For the scattering formulation, we consider the schematics in Fig 11.
We denote all the basis state incoming to the site $(j, k)$ (continuous 
arrows in Fig. 11) by $| \sigma, (j, k) \rangle$.
Since now the lattice coordination is 3, we need more than the usual $\pm$ 
to label the direction quantum number $\sigma$.
Therefore, we assume that $\sigma$ has the values $0, 1, 2$, in analogy 
with the coined model.
But in each bond we can ascribe only two of these possible values to the 
two states representing opposite directions.
In this way, for the general situation represented in Fig. 11, we set the 
convention (which can be checked to be self-consistent along the whole 
lattice)

\begin{equation}
\sigma_a = [j]_3, \ \ \sigma_b = [j+1]_3, \ \ 
\sigma_c = [j-1]_3.
\label{sigma-convention}
\end{equation}

At each bond, for a state $|\sigma, (j, k) \rangle$ incoming to $(j,k)$ 
there is a corresponding outgoing state, which is itself an incoming state 
to another site $(j', k')$ and denoted by $|\sigma' , (j', k')\rangle$
(dashed arrows in Fig. 11).
From the protocol used to label the $\sigma$'s -- Fig. 11 and 
Eq. (\ref{sigma-convention}) -- it is not difficult to show that 
$\sigma' = \phi_k(\sigma)$, $j' = f(j, k; \phi_k(\sigma))$, and
$k' = g(j, k; \phi_k(\sigma))$, for $\phi$, $f$ and $g$ given in Eq. 
(\ref{f-g}).
For instance, in Fig. 11 for $k$ even and $[j]_3 = 0$, $\sigma_a = 0$ 
and the other state in the same bond than $|0, (j, k) \rangle$ is 
$|2, (j+1, k-1) \rangle$, as verified by direct inspection.
Likewise, this follows directly from
$\sigma' = \phi_{k-{\mbox{\scriptsize even}}}(0) = 2$,
$j' = f(j, k; 2)|_{k-{\mbox{\scriptsize even}},[j]_3 = 0} = j+1$ and 
$k' = g(j, k; 2))|_{k-{\mbox{\scriptsize even}},[j]_3 = 0} = k-1$.

Thus, the evolution operator is $U_{s} = T + R$, with  
\begin{eqnarray}
T |\sigma, (j, k) \rangle 
&=& \sum_{\alpha \neq \sigma}
t_{\phi_k(\alpha),  \, \sigma}^{(j,k)}  \,  
| \phi_k(\alpha), (f(j,k;\phi_k(\alpha)), g(j,k;\phi_k(\alpha))) \rangle,
\nonumber \\ 
R |\sigma, (j, k) \rangle 
&=&
r_{\phi_k(\sigma),  \, \sigma}^{(j,k)}  \,
|\phi_k(\sigma), (f(j,k;\phi_k(\sigma)), g(j,k;\phi_k(\sigma))) \rangle,
\label{r-t-honeycomb}
\end{eqnarray}
where furthermore
\begin{eqnarray}
\hspace{-2cm} T^{\dagger} |\sigma, (j, k) \rangle 
&=& \sum_{\alpha \neq \phi_k(\sigma)}
[t_{\sigma, \, \alpha}^{(f(j,k;\phi_k(\sigma)), g(j,k;\phi_k(\sigma)))}]^{*}  \, 
| \alpha, (f(j,k;\phi_k(\sigma)), g(j,k;\phi_k(\sigma))) \rangle,
\nonumber \\ 
\hspace{-2cm} R^{\dagger} |\sigma, (j, k) \rangle &=&
[r_{\sigma, \, \phi_k(\sigma)}^{(f(j,k;\phi_k(\sigma)), g(j,k;\phi_k(\sigma)))}]^{*}  
\, |\phi_k(\sigma), (f(j,k;\phi_k(\sigma)), g(j,k;\phi_k(\sigma))) \rangle.
\label{r-t-star-honeycomb}
\end{eqnarray}

Now, if we define
\begin{equation}
\Gamma^{(j, k)} =  
\left(
    \begin{array}{ccc}
\Gamma^{(j, k)}_{0 \, 0} & \Gamma^{(j, k)}_{0 \, 1} & \Gamma^{(j, k)}_{0 \, 2} \\
\Gamma^{(j, k)}_{1 \, 0} & \Gamma^{(j, k)}_{1 \, 1} & \Gamma^{(j, k)}_{1 \, 2} \\
\Gamma^{(j, k)}_{2 \, 0} & \Gamma^{(j, k)}_{2 \, 1} & \Gamma^{(j, k)}_{2 \, 2}  
    \end{array}
\right),
\label{scattering-tres}
\end{equation}
and identify ($\alpha, \beta = 0, 1, 2$)
\begin{equation}
r_{\phi_k(\sigma), \,  \sigma} = \Gamma^{(j, k)}_{\phi_k(\sigma) \,  \sigma},
\ \ \ 
t_{\beta, \,  \alpha} = \Gamma^{(j, k)}_{\beta \, \alpha} \ 
(\mbox{for} \ \beta \neq \phi_k(\alpha)), 
\label{relation-rtgama}
\end{equation}
one has that the unitarity of $\Gamma$ in Eq. 
(\ref{scattering-tres}) guarantees that $U_{s}$
is also unitary.

Finally, the projector operator reads
\begin{eqnarray}
\hspace{-2cm} P_{s}^{(j, k)} = & &  
|\phi_k(\sigma), (f(j,k;\phi_k(\sigma)), g(j,k;\phi_k(\sigma)))
\rangle \langle 
(f(j,k;\phi_k(\sigma)), g(j,k;\phi_k(\sigma))), \phi_k(\sigma) |
\nonumber \\
\hspace{-2cm} & & + |\sigma, (j, k) \rangle \langle (j, k), \sigma |.
\label{projector-honeycomb-scattering}
\end{eqnarray}

\subsubsection{Mapping the models} \

First, we note that: (i) comparing the $\sigma$ labeling convention for
the coined and scattering formulation in the schematics in Fig. 11 (see
also Eq. (\ref{sigma-convention})); and (ii) taking into account how the 
states evolve according to the quantum number $\sigma$ in both 
models; it turns out that a direct one-to-one association between 
basis states is simply given by 
\begin{equation}
E |\sigma, (j, k) \rangle = 
|(j, k) \rangle \otimes | \sigma \rangle .
\label{e-honeycomb}
\end{equation}

Second, if as done in Sec. 3.1, we set the coin matrix Eq. (\ref{coin-tres}) 
and the scattering coefficients in Eq. (\ref{scattering-tres}) to be equal, 
then we find that the corresponding expressions for $U_s$ and $U_c$, 
with $E$ given by Eq. (\ref{e-honeycomb}), satisfies to Eq. 
(\ref{unitary-equivalence-1d}).

Lastly, to obtain the probabilities of one model by means of the other, we 
define the cross operators as in Eq. (\ref{projections-mapping-1d}), 
using the definitions in Eqs. (\ref{projector-honeycomb-coin}), 
(\ref{projector-honeycomb-scattering}) and (\ref{e-honeycomb}).

\subsubsection{Examples} \

To illustrate the dynamics in a honeycomb topology, we again consider 
different coin (and equivalent scattering) matrices and calculate 
the QWs time evolutions. 
We analyze the following five different examples for the transition
probabilities (assumed to be the same at all the lattice sites): \\
\\
(a) $3 \times 3$ typical unbiased operator
\begin{equation}
C_{unb_3} = 
\frac{1}{\sqrt{3}}
\left(
\begin{array}{ccc}
      -1 & \exp[- \pi i / 3] & \exp[- \pi i / 3] \\
       \exp[- \pi i / 3] & -1 & \exp[- \pi i / 3] \\
       \exp[-\pi i / 3] & \exp[-\pi i / 3] & -1 
\end{array}
\right); 
\label{honey-arb-complex}
\end{equation}
(b)  $3 \times 3$ biased (and real) operator
\begin{equation}
C_{bia_3} = \frac{1}{3}
\left(
\begin{array}{ccc}
      1 \ \ & 1 - \sqrt{3} \ \ & 1 + \sqrt{3} \\
      1  + \sqrt{3} \ \ & 1 \ \ & 1 - \sqrt{3} \\
      1 - \sqrt{3} \ \ & 1 + \sqrt{3} \ \ & 1
\end{array}
\right);
\label{honey-arb-real} 
\end{equation}
(c) $3 \times 3$ Discrete Hartley Transform (DHT) \cite{ersoy}
\begin{equation}
C_{DHT_3} = \frac{1}{2 \sqrt{3}}
\left(
\begin{array}{ccc}
      2 \ \ & 2 \ \ & 2 \\
      2 \ \ & -1 + \sqrt{3} \ \ & -1 - \sqrt{3} \\
      2 \ \ & -1- \sqrt{3} \ \ & -1 + \sqrt{3}
\end{array}
\right);
\label{honey-dht} 
\end{equation}
(d) $3 \times 3$ Grover
\begin{equation}
C_{G_3} = 
\frac{1}{3}
\left(
\begin{array}{ccc}
      -1 &  2 & 2 \\
       2 & -1 & 2 \\
       2 &  2 & -1
\end{array}
\right);
\label{honey-grover}
\end{equation}
(e) $3 \times 3$ Discrete Fourier Transform (DFT)
\begin{equation}
C_{DFT_3} = 
\frac{1}{\sqrt{3}}
\left(
\begin{array}{ccc}
      \exp[2 \pi i / 3] & 1 & \exp[- 2 \pi i / 3] \\
      1 & 1 & 1 \\
      \exp[-2 \pi i / 3] & 1 & \exp[2 \pi i / 3]
\end{array}
\right).
\label{honey-dft}
\end{equation}

\begin{figure}
\centerline{\psfig{figure=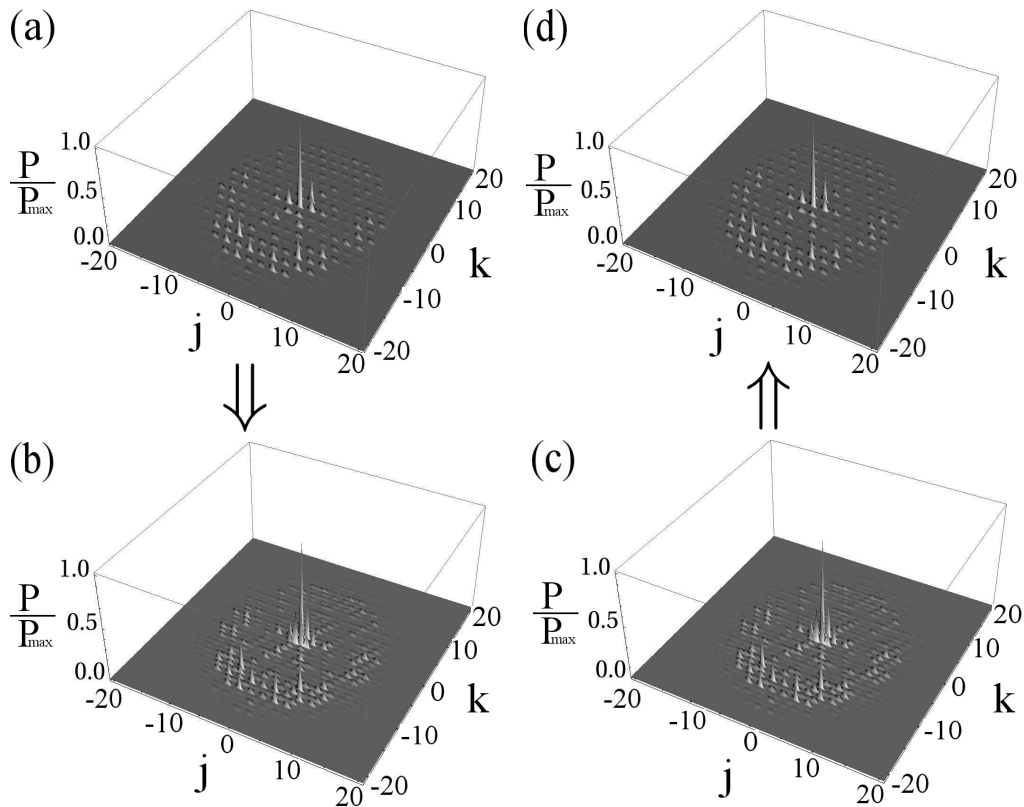,width=13.5cm}}
\caption{Quantum walks in the  honeycomb lattice considering the 
$3 \times 3$ typical unbiased transition coefficients, Eq. 
(\ref{honey-arb-complex}).
The density plots represent the probabilities to be in the different 
states (as defined by the projector operators) after $n=20$ time steps 
for $|\Psi(0)\rangle$ given in the main text.
The results are for the: (a) coined, (b) scattering obtained from the 
coined, (c) scattering, and (d) coined obtained form the scattering,
models.
The graphs (a) and (d) and (b) and (c) are complete identical.}
\end{figure}

Few comments about the matrices above are in order.
The first case, Eq. (\ref{honey-arb-complex}), is a typical (complex)
unbiased operator since the coefficients have the same value for 
their modulus square, $|c_{\sigma' \, \sigma}|^2 = 1/3$.
Therefore, the resulting transition probabilities are equal to 1/3.
On the other hand, the second, Eq. (\ref{honey-arb-real}) -- here with
all entries real numbers -- is strongly biased because the transition 
probabilities are very different (but obviously summing up to 1).
For instance, for the scattering formulation, $k$-even and $[j]_3 = 0$
in Fig. 11 (cf. Eq. (\ref{relation-rtgama})), we have 
$|t_{0, 0}|^2 = 1/9 \approx 0.11$, 
$|t_{1, 0}|^2 = 2 (2+\sqrt{3})/9 \approx 0.83$, 
$|r_{2, 0}|^2 = 2 (2-\sqrt{3})/9 \approx 0.06$.
The discrete Hartley transform matrix \cite{ersoy}, Eq. (\ref{honey-dht}), 
although not usually considered in QWs, is an interesting example to study
due to its usefulness in signal processing \cite{boussakta}, furthermore 
always being real. 
The other two are just the $3 \times 3$ versions of the Grover and DFT.
Finally, note there is not a $3 \times 3$ Hadamard matrix.

Assuming as the initial states
\begin{equation}
| \Psi(0) \rangle_s = |1, \, (0,0) \rangle, \qquad
E \, | \Psi(0) \rangle_s =  | \Psi(0) \rangle_c = |(0,0) \rangle \otimes 
| 1 \rangle, 
\label{initial-honeycomb}
\end{equation}
we evolve the models 20 time steps.
The resulting probability patterns are displayed in Figs. 12--16.
The graphs are organized exactly as done in the Sec. 3.1.4 (Figs. 5--8).

\begin{figure}[top]
\centerline{\psfig{figure=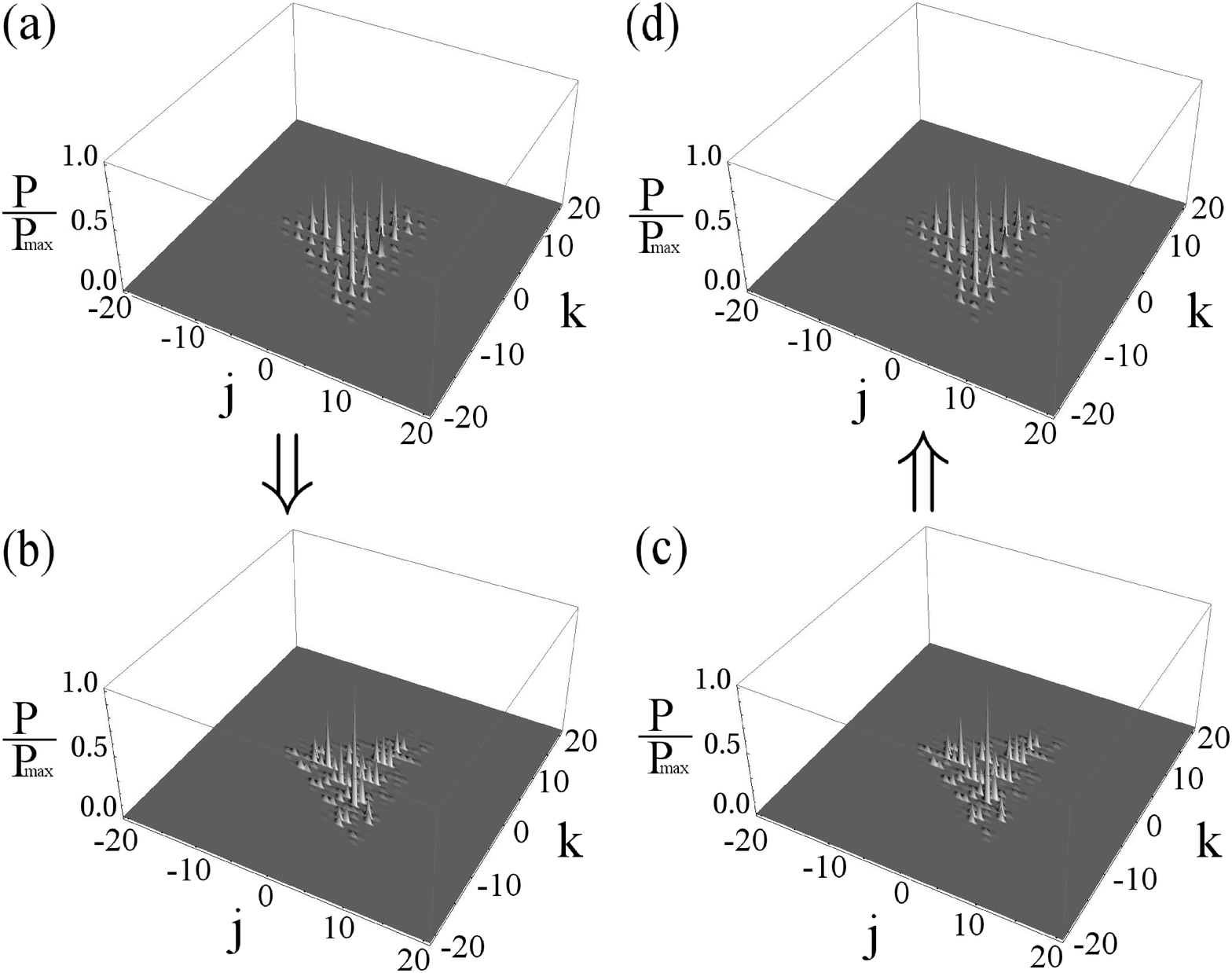,width=13.5cm}}
\caption{The same as in Fig. 12, but for the $3 \times 3$ biased real
matrix, Eq. (\ref{honey-arb-real}).}
\end{figure}

\begin{figure}[top]
\centerline{\psfig{figure=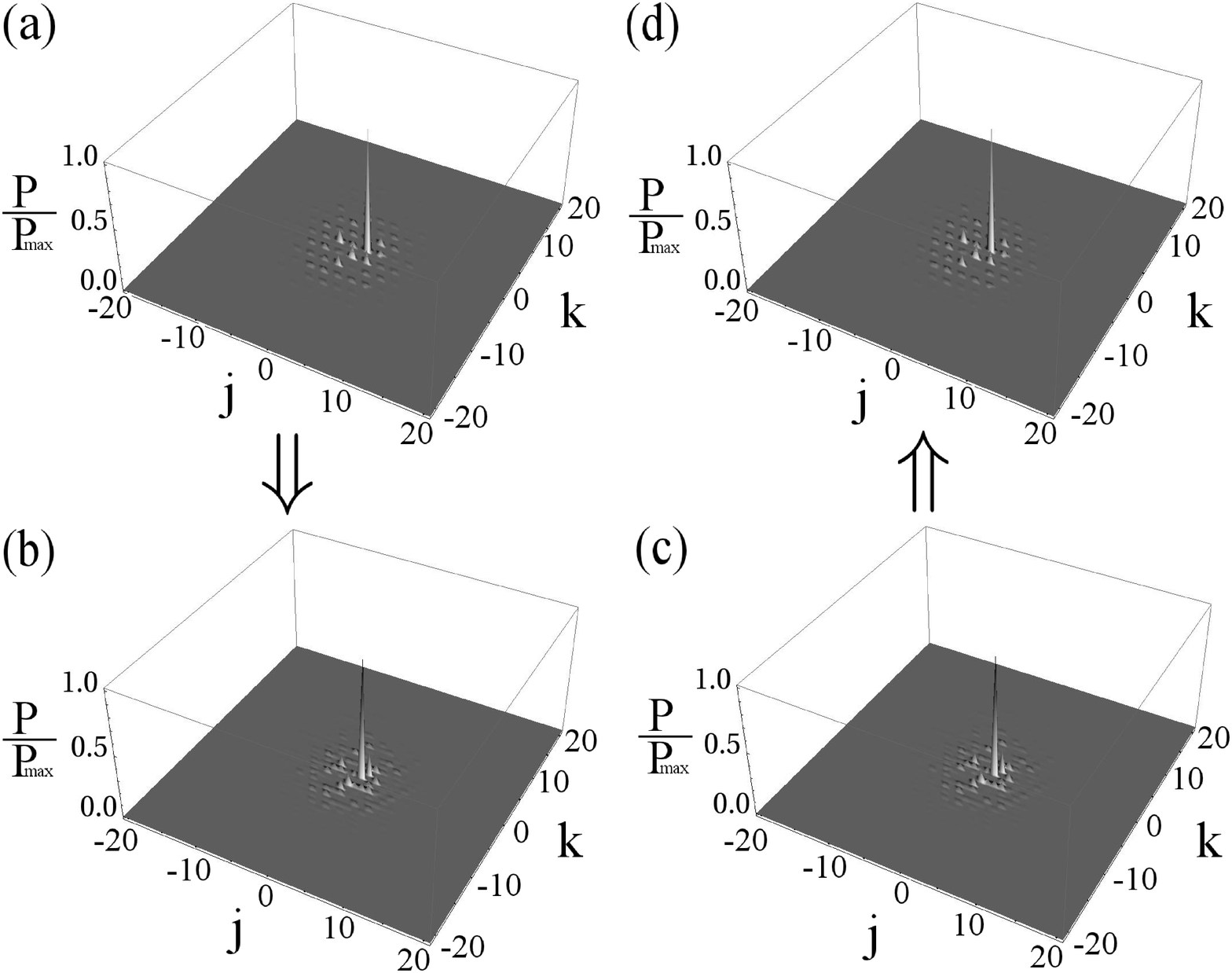,width=13.5cm}}
\caption{The same as in Fig. 12, but for the $3 \times 3$ DHT, 
Eq. (\ref{honey-dht}).}
\end{figure}

\begin{figure}[top]
\centerline{\psfig{figure=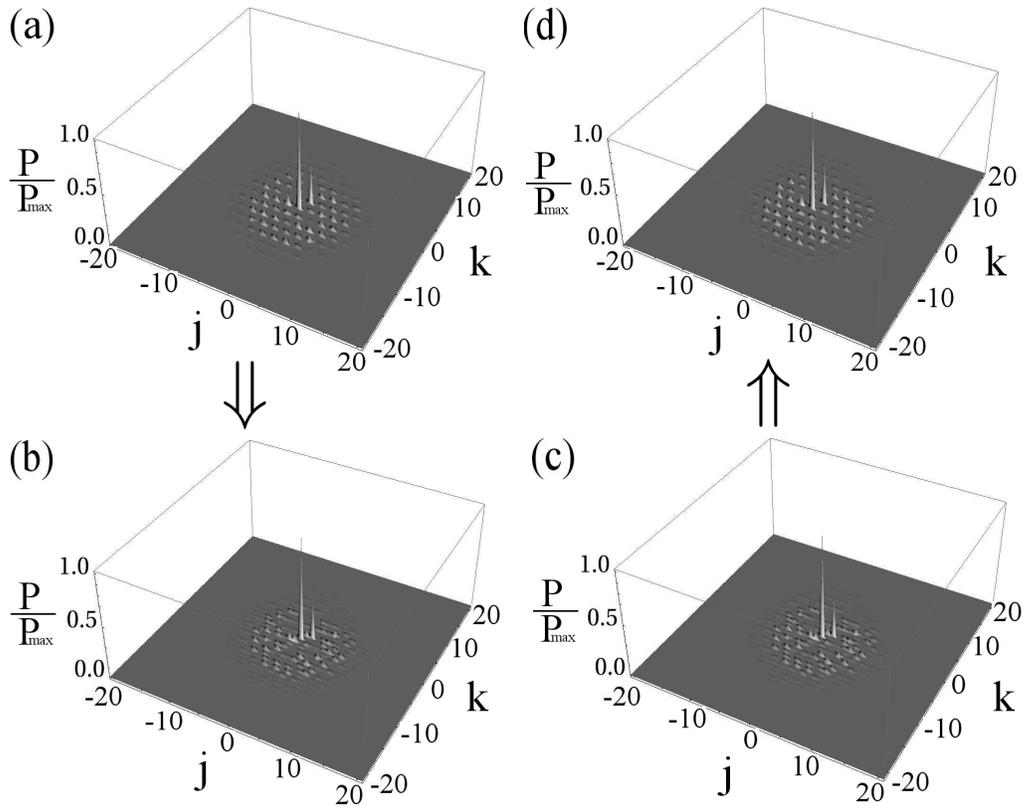,width=13.5cm}}
\caption{The same as in Fig. 12, but for the $3 \times 3$ Grover, 
Eq. (\ref{honey-grover}).}
\end{figure}

\begin{figure}[top]
\centerline{\psfig{figure=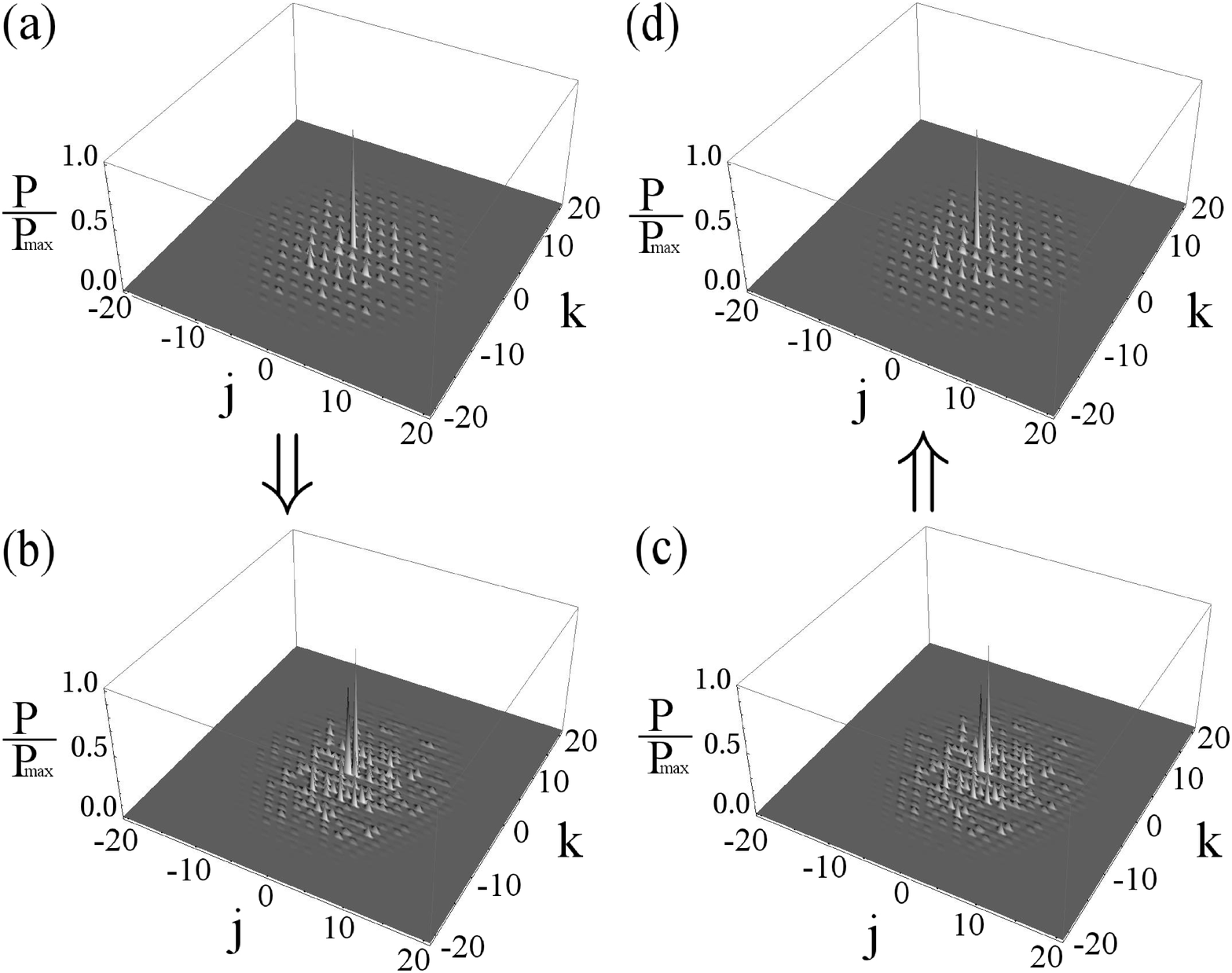,width=13.5cm}}
\caption{The same as in Fig. 12, but for the $3 \times 3$ DFT, 
Eq. (\ref{honey-dft}).}
\end{figure}

Again, certain facts are clearly observed from Figs. 12--16.
First, as it should be, we can obtain one evolution from the other
by correct projections.
Second, like in the square lattice, in the honeycomb the CQW
probabilities are in general more sparse and somewhat smoother than those 
for the SQW.
And third, it is interesting to notice the particular pattern in Fig. 13, 
with a tendency of three preferable directions of propagation along the 
lattice.
This is a consequence of the probabilities bias resulting from the matrix 
$C_{bia_3}$, Eq. (\ref{honey-arb-real}), for which three transition elements 
$c_{0 \, 2} = c_{1 \, 0} = c_{2 \, 1} = (1+\sqrt{3})/3$ are considerable higher 
than the other six.

Finally, comparing the Grover (DFT) probability amplitudes for the square 
and honeycomb topologies, respectively, Figs. 7 and 15 (Figs. 8 and 16)
we can have an idea on the influence of a regular lattice coordination 
number to the QW dynamics.
In fact, note that the $|\Psi(0)\rangle$ used for the square lattice 
examples is equally ``distributed'' among the four $\sigma$ values basis 
states, Eq. (\ref{initial-square}), which is not the case for the honeycomb 
lattice whose $|\Psi(0)\rangle$ is written in terms of just one value of 
$\sigma$ (from the three possible), Eq. (\ref{initial-honeycomb}).
Even then, we see that in the latter the resulting probabilities are more 
uniformly distributed, specially in the Grover case.
This illustrates the intricate process of quantum transitions at each 
lattice site as function of the number of bonds attached to it.

\section{Conclusion}

By means of detailed analysis of specific examples, in this work we have
compared the different aspects associated to the construction of the 
discrete time coined and scattering quantum walks.
Specifically, we have presented a complete formulation for the line, 
square and the involving honeycomb lattice topologies.
In all the examples studied, we have clarified how to map one model version 
to the other.
Also, we have illustrated the particularities in the probability pattern
distributions along the graph structures, resulting from the different time 
evolution of each formulation.
Finally, distinct coin (and corresponding scattering) matrices operators 
were considered. 
Besides the usual Hadamard, Grover and Discrete Fourier Transform, we have 
addressed the Discrete Hartley Transform as well as a few other cases.

Our aim here was to show through concrete situations that in fact the CQWs 
and SQWs are closely related.
Moreover, the use of a specific QW version in different applications may be 
more a matter of practicality (or even of taste, as recently stressed
\cite{hillery-search-1}), than due to prohibitive conceptual differences 
between the models.

Regarding implementations, the issue may be a little subtle considering
that usually distinct experimental architectures are proposed for CQWs and 
for SQWs (see the Introduction Sec.).
However, a given realization of a specific QW model conceivably could 
also be  used to obtain results from the other model. 
But then, the experimental setup somehow should allow the effective
construction of proper cross projectors (as defined in Secs. 2-3 and 2-4).
The important point is that efforts in this direction 
certainly would pay off since an actual implementation of one version of 
QWs having proper built-in projections to the other version -- 
thus ``carrying'' the features of both type of QWs -- would be much more 
flexible and useful in applications.

\section*{Acknowledgements}

We are in great debt to M. C. Santos for help with the numerics.
Research grants are provided by CNPq (da Luz) and Finepe-CTInfra.


\section*{References}

\begin{thebibliography}{mt1}


\bibitem{Phys.Rev.A.48.1687.1993}
Aharonov Y, Davidovich L, and Zagury N
1993
{\em Phys. Rev.} A {\bf 48}
1687

\bibitem{jsp.1996.85.551}
Meyer D A
1996
{\em J. Stat. Phys.} {\bf 85}
551

\bibitem{watrous-2001}
Watrous S
2001
{\em J. Comput. Sys. Sci.} {\bf 62}
376

\bibitem{kempe1}
Kempe J
2003
{\em Contemp. Phys.} {\bf 44}
307

\bibitem{venegas}
Venegas-Andraca S E
2012
{\em Quantum Information Processing} {\bf 11}
1015

\bibitem{ctv}
Farhi E and Gutmann S
1998
{\em Phys. Rev.} A {\bf 58} 
915 \\
Childs A M, Farhi E, and Gutmann S
2002
{\em Quant. Info. Proc.} {\bf 1}
35 \\
Childs A M, Cleve R, Deotto E, Farhi E, Gutmann S, and Spielman D A
2003
in {\em Proceedings of the 35th ACM Symposium on Theory of Computing
(STOC 2003)}
(ACM Press, New York), pp. 59-68

\bibitem{tregenna1}
Tregenna B, Flanagan W, Maile R, and Kendon V
2003
{\em New J. Phys.} {\bf 5}
83

\bibitem{hillery-2003}
Hillery M, Bergou J, and Feldman E 
2003
{\em Phys. Rev.} A {\bf 68}
032314

\bibitem{hillery-2004}
Feldman E and Hillery M
2004
{\em Phys. Lett.} A {\bf 324}
277

\bibitem{hillery-2007}
Feldman E and Hillery M
2007
{\em J. Phys.} A {\bf 40}
11343

\bibitem{search}
Shenvi S, Kempe J, Whaley K B
2003
{\em Phys. Rev.} A {\bf 67}
052307 \\
Childs A M and Goldstone J
2004
{\em Phys. Rev.} A {\bf 70}
022314 

\bibitem{search1}
Gabris A, Kiss T, and Jex I
2007
{\em Phys. Rev.} A {\bf 76}
062315

\bibitem{hillery-search}
Hillery M, Reitzner D, and Buzek V
2010
{\em Phys. Rev.} A {\bf 81}
062324

\bibitem{hillery-search-1}
Hillery M, Zheng H, Feldman E, Reitzner D, and Buzek V
2012
{\em Phys. Rev.} A {\bf 85}
062325

\bibitem{kosik}
Kosik J and Busek V
2005
{\em Phys. Rev.} A {\bf 71}
012306

\bibitem{childs1}
Childs A M
2009
{\em Phys. Rev. Lett.} {\bf 102}
180501

\bibitem{karski}
Karski M, Foerster L, Choi J-M, Steffen A, Alt W, Meschede D, 
and Widera A
2009
{\em Science} {\bf 325}
174

\bibitem{leung}
Leung G, Knott P, Bailey J, and Kendon V
2010
{\em New J. Phys.} {\bf 12}
123018

\bibitem{kempe-ptrf}
Kempe J
2005
{\em Probab. Theory Relat. Fields} {\bf 133}
215

\bibitem{biology}
Karafyllidis I G and Lagoudas D C
2007
{\em Biosystems} {\bf 88}
137 

\bibitem{biology1}
Mohseni M, Rebentrost P, Loyd S, and Aspuru-Guzik A
2008
{\em J. Chem. Phys.} {\bf 129}
174106 \\
Rebentrost P, Mohseni M, Kassal I, Loyd S, and Aspuru-Guzik A
2009
{\em New J. Phys.} {\bf 11}
033003 

\bibitem{chandrashekar1}
Chandrashekar C M
2011
{\em Phys. Rev.} A {\bf 83}
022320

\bibitem{chandrashekar2}
Chandrashekar C M and Laflamme R
2008
{\em Phys. Rev. A} {\bf 78}
022314

\bibitem{ampadu}
Ampadu A
2012
{\em Commum. Theor. Phys.} {\bf 57}
41

\bibitem{game}
Chandrashekar C M and Banerjee S
2011
{\em Phys. Lett.} A {\bf 375}
1553

\bibitem{qw-qcomput}
Aharonov D, Ambainis A, Kempe J, and Vazirani U
2001 
in {\em STOC'01: Proceedings of the 33nd Annual ACM Symposium on Theory 
of Computing} 
(ACM, New York), pp. 37--49 \\ 
Ambainis A, Bach E, Nayak A, Vishwanath A, and Watrous J 
2001
in {\em STOC'01: Proceedings of the 33nd Annual ACM
Symposium on Theory of Computing} 
(ACM, New York), pp. 50--59
\\
Lovett N B, Cooper S, Everitt M, Trevers M, and Kendon V, 
2010
{\em Phys. Rev.} A {\bf 81} 
042330
\\
Ambainis A
2008 
in {\em SOFSEM 2008: Theory and Practice of Computer Science}, 
edited by Geffert V, Karhumaki J, Bertoni A, Preneel B, Navrat P, 
and Bielikova M
(Springer, Berlin), pp. 1--4
\\
Mosca M
2009
in {\em Quantum Algorithms, Encyclopedia of Complexity Systems Science}, 
edited by Meyers R A 
(Springer, Heidelberg)

\bibitem{dur}
Dur W, Raussendorf R, Kendon V M, and Briegel H J
2002
{\em Phys. Rev.} A {\bf 66}
052319

\bibitem{hoogdalem}
van Hoogdalem K A and Blaauboer M
2009
{\em Phys. Rev.} B {\bf 80}
125309

\bibitem{zou}
Zou X, Dong Y, and Guo G 
2006
{\em New J. Phys.} {\bf 8}
81

\bibitem{tiegang}
Tiegang D, Hillery M, and Zubairy M S
2004
{\em Phys. Rev.} A {\bf 70}
032304

\bibitem{crespi}
Crespi A, Sansoni L, Vallone G, Sciarrino F, Ramponi R,
Mataloni P, and Osellame R
2012
in {\em Frontiers in Ultrafast optics: biomedical, scientific,
and industrial applications XII} edited by 
Heisterkamp A, Meunier M, and Nolte S, 
Proc. of SPIE {\bf 8247}, 82470L
(Washington: SPIE Press)

\bibitem{perets}
Perets H B, Lahini Y, Pozzi F, Sorel M, Morandotii R, and Silberberg Y
2008
{\em Phys. Rev. Lett.} {\bf 100}
170506

\bibitem{schreiber}
Schreiber A, Cassemiro K N, Potocek V, Gabris A, Mosley P J, Andersson E,
Jex I, and Silberhorn Ch
2010
{\em Phys. Rev. Lett.} {\bf 104}
050502

\bibitem{ryan}
Ryan C A, Laforest M, Boileau J C, and Laflamme R
2005
{\em Phys. Rev.} A {\bf 72}
062317

\bibitem{matjeschk}
Matjeschk R, Schneider CH, Enderlein M, Huber T,
Schmitz H, Glueckert J, and Schaetz T
2012
{\em New J. Phys.} {\bf 14}
035012

\bibitem{jex-beam}
Jex I, Stenholm S, and Zeilinger A
1995
{\em Opt. Commun.} {\bf 117}
95

\bibitem{torma}
Torma P and Jex I
1999
{\em J. Opt.} B {\bf 1}
8

\bibitem{do}
Do B, Stohler M L, Balasubramanian S, Elliott D S, Eash C, Fischbach E, 
Fischbach M A, Mills A, and Zwickl B
2005
{\em J. Opt. Soc. Am.} B {\bf 22}
499

\bibitem{strauch}
Strauch F W
2006
{\em Phys. Rev.} A {\bf 74}
030301(R) 

\bibitem{childs-limit}
Childs A M
2010
{\em Commun. Math. Phys.} {\bf 294}
581

\bibitem{andrade1}
Andrade F M and da Luz M G E 
2009
{\em Phys. Rev.} A {\bf 80} 
052301

\bibitem{andrade-2011}
Andrade F M and da Luz M G E
2011
{\em Phys. Rev.} A {\bf 84}
042343

\bibitem{ambainis-2003}
Ambainis A
2003
{\em Int. J. Quant. Inform.} {\bf 1}
507

\bibitem{prl.2004.93.190503}
Ribeiro P, Milman P, and Mosseri R
2004
{\em Phys. Rev. Lett.} {\bf 93}
190503

\bibitem{nielsen}
Nielsen M A and Chuang I L
2000
{\em Quantum computation and quantum information}
(Cambridge: Cambridge Univ. Press)

\bibitem{bach-2004}
Bach E, Coppersmith S, Goldschen M P, Joynt R, and Watrous J
2004
{\em J. Comput. Syst. Sci.} {\bf 69}
562

\bibitem{carneiro-2005}
Carneiro I, Loo M, Xu X, Girerd M, Kendon V, and Knight P L
2005
{\em New J. Phys.} {\bf 7}
156

\bibitem{pra.2008.77.032326}
Chandrashekar C M, Srikanth R, and Laflamme R
2008
{\em Phys. Rev.} A {\bf 77}, 
032326

\bibitem{schmidt1}
da Luz M G E, Cheng B K, and Heller E J
1998
{\em J. Phys.} A {\bf 31}
2975 \\
Schmidt A G M, Cheng B K, and da Luz M G E
2002
{\em Phys. Rev.} A {\bf 66}
062712 \\
Schmidt A G M, Cheng B K, and da Luz M G E
2003
{\em J. Phys. A} {\bf 36}
L545 \\
Zanetti F M, Kuhn J, Delben G J, Cheng B K, and da Luz M G E
2006
{\em J. Phys. A} {\bf 39}
2493

\bibitem{chadan}
Chadan K and Sabatier P C
1989
{\em Inverse problems in quantum scattering theory} 2nd. edition
(Berlin: Springer-Verlag)

\bibitem{andrade-interferencia}
Andrade F M and da Luz M G E
2012
{\em Phys. Rev.} A {\bf 86}
042309

\bibitem{mackay}
Mackay T D, Bartlett S D, Stephenson L T, and B C Sanders
2002
{\em J. Phys.} A {\bf 35}
2745

\bibitem{carneiro}
Carneiro I, Loo M, Xu X, Girerd M, Kendon V, and Knight P L
2005
{\em New J. Phys.} {\bf 7}
156

\bibitem{diagonal}
Oliveira A C, Portugal R, and Donangelo R
2006
{\em Phys. Rev.} A {\bf 74}
012312

\bibitem{inui}
Inui N, Konno N, and Segawa E
2005
{\em Phys. Rev.} E {\bf 72}
056112

\bibitem{jafarizadeh}
Jafarizadeh M A and Sufiani R
2007
{\em Physica} A {\bf 381}
116

\bibitem{abal}
Abal G, Donangelo R, Marquezinho F L, and Portugal R
2010
{\em Math. Struct. Comput. Sci.} {\bf 20}
999

\bibitem{kollar}
Kollar B, Stefanak M, Kiss T, and Jex I
2010
{\em Phys. Rev.} A {\bf 82}
012303

\bibitem{ersoy}
Ersoy O K
1994
{\em Proceedings IEEE} {\bf 82}
429

\bibitem{boussakta}
Boussakta S, Alshibami O H, and Aziz M Y
2001
{\em IEEE Trans. Signal Process.} {\bf 49}
3145

\end{thebibliography}
\end{document}